\documentclass[aps, prd, amsmath, floats, floatfix, onecolumn, nofootinbib]{revtex4-2}
\usepackage{color} 
\usepackage{mathrsfs}
\usepackage{amsmath}
\usepackage{graphicx}
\usepackage[hidelinks]{hyperref}
\usepackage{tabularx} 
\usepackage{float} 
\newcommand{\be}{\begin{equation}}
\newcommand{\ee}{\end{equation}}
\newcommand\bea{\begin{eqnarray}}
\newcommand\eea{\end{eqnarray}}
\newcommand{\dd}{\mbox{d}}
\newcommand{\sn}{\mbox{s}}
\newcommand{\cn}{\mbox{c}}

\begin{document}

\author{Zhongyou Mo}
\email[]{11930796@mail.sustech.edu.cn}
\author{Leonardo Modesto}
\email[]{lmodesto@sustech.edu.cn}
\affiliation{Department of Physics, Southern University of Science and Technology, Shenzhen, 518055, China}

\title{Quantum interference in the Kerr spacetime}
\date{today}

\begin{abstract}
The gravitational induced interference is here studied in the framework of Teleparallel Gravity. We derive the gravitational phase difference and we apply the result to the case of a Kerr spacetime. Afterwards, we compute the fringe shifts in an interference experiment of particles and discuss how to increase their values by changing the given parameters that include: the area in between the paths, the energy of the particles, the distance from the black hole, the mass and the spin of the black hole. It turns out that it is more difficult to detect the fringe shifts for massless particles than for massive particles. As a further application, we show how the mass of the black hole and its angular momentum can be obtained from the measurement of the fringe shifts. Finally, we compare the phase difference derived in Teleparallel Gravity with a previous work in General Relativity.

\end{abstract}

\maketitle

\tableofcontents

\section{Introduction\label{Intro}}

In the year 1959, Aharonov and Bohm proposed an observable effect due to electromagnetic potentials in the quantum domain~\cite{Aharonov:1959fk}. 
They showed that, contrary to the conclusions of classical mechanics, in quantum mechanics there are effects of electromagnetic potentials on charged particles, even in the region where all the fields vanish. 
 In their model, two electron beams go through two cylindrical tubes within two different time-dependent potentials, to finally interfere in a region outside the tubes. 
 In particular, they proved that the interference depends on the time integrals of the potentials. 
The same two authors proposed also another experiment that we summarize as follows. In the region outside an infinite cylindrical solenoid (in which a magnetic field is confined), an electron beam is split in two, one travels to the right while the other to the left of the solenoid and then they interfere. It turns out that the interference of the two beams depends on the contour integral of the vector potential. 
These thought experiments prove that even in  regions where the fields are absent, the electromagnetic potential can affect the observations.

To clarity, we write the wave function in the presence of the potential as~\cite{Aharonov:1959fk}
\be
\Psi={\Psi_1}^0 e^{-\frac{i}{\hbar} \mathscr{S}_1 }
+{\Psi_2}^0 e^{-\frac{i}{\hbar} \mathscr{S}_2 },
\label{Psi function}
\ee
where ${\Psi_1}^0$ and ${\Psi_2}^0$ denote the free wave functions. It turns out that the interference depends on the difference between of the two phase factors in \eqref{Psi function}. In general, the phase difference is given by~\cite{Wu:1975es}
\be
-\frac{1}{\hbar}(\mathscr{S}_1-\mathscr{S}_2)=\frac{e}{\hbar c} \oint A_\mu \dd x^\mu,
\label{phase difference electron}
\ee
where the closed integral is unshrinkable. In the second thought experiment mentioned above, the right hand side of \eqref{phase difference electron} is proportional to the magnetic flux through the cross section of the solenoid. The Aharonov-Bohm effect (AB effect) caused by a magnetic field was experimentally observed by Chambers~\cite{Chambers:1960xlk}. Since then, more observations for the AB effect were performed (see Ref.~\cite{Tonomura:1989zz} for a review of them).

As a route to connect general relativity with quantum mechanics, it is appealing to image a phase induced by the gravitational field in analogy with the one by the electromagnetic field. The effect of the gravity induced phase, analogous to the AB effect in electromagnetism, is usually referred to as Gravitational Aharonov-Bohm effect~\cite{Dowker,Ford,Audretsch:1982ux}. In gravity, the interference of particles moving in a flat spacetime region may be affected by a non vanishing Riemann tensor localized far from the particles. In Ref.~\cite{Stodolsky:1978ks}, Stodolsky argued that such phase is given by 
\be 
\frac{mc}{\hbar} \int \dd s 
\ee
for a massive particle (in the case of a semiclassical limit in which particles travel along the classical path).
An interesting feature of this expression is its property under coordinate transformations. As Stodolsky showed, the above phase is gauge invariant under coordinate transformations, as opposite to the gauge variance of the  electromagnetic phase under $U(1)$ transformations of the potential. 
This discovery reveals the difference between the symmetry properties of the gravitational and the electromagnetic field in the quantum domain. 

Concerning our work, we will evaluate the phase in the theory of Teleparallel Gravity (TG). This theory is also known as the Teleparallel Equivalent of General Relativity~\cite{Aldrovandi:2013wha}. In TG, the phase $(mc/\hbar)\int \dd s$ can be separated into three parts~\cite{Aldrovandi:2013wha}: the first part represents the free particle, while the second part stands for the inertial effects of the frame, which can be eliminated by choosing an inertial frame, the third part is the one we really have to take care of. Indeed, it represents the gravitational interaction given by the integral of a gauge potential for gravity. Our study is based on this formulation.

Before getting to the heart of our contribution, it deserves to be mentioned the experimental work on the gravitational phase. In 1974, Overhauser and Colella proposed an experiment to detect the gravitational quantum interference~\cite{Overhauser:1974}. In their proposal, a neutrons' beam is split into two parts and recombined afterwords.
The trajectories of the neutrons approximately form a vertical parallelogram with its base parallel to the surface of the earth. 
They found that the phase difference between the two beams is related to the gravitational acceleration. 
In the next year, Colella  {\em et al.} implemented such idea experimentally~\cite{Colella:1975dq}. They rotated the interferometer to change the angle between the parallelogram and the surface of the earth, and detected the corresponding counting rates of the interfering beams. With these results they determined the number of the fringes caused by the gravity. Although the influence of the gravitational field of the earth has been found, the gravitational interference caused by small masses is still a difficult task. 
On this subject, Hohensee {\em et al.} proposed an experiment in which matter waves are in a gravitational potential of a pair of masses with vanishing net gravitational force~\cite{Hohensee:2011yt}. This thought experiment has not been realized because it requires the optical lattice to be perfect (see the comment in Ref.~\cite{Roura:2021fvd}). 
Recently the gravitational interference caused by small masses has been detected by Overstreet {\em et al.} experimentally~\cite{Overstreet:2021hea}
\footnote{In Ref.~\cite{Overstreet:2021hea} the authors claim they have observed the gravitational Aharonov-Bohm effect. Such result is extremely interesting, but we should notice that the observed effect is not exactly the one in Refs.~\cite{Dowker,Ford,Audretsch:1982ux} because the atoms move in a region where the Riemann curvature does not vanish.}, using laser pulses to split and recombine two atoms vertically at different times. The upper atom goes closer to a ring mass than the lower atom, which leads to a gravity induced phase difference between these atoms. 

Now that the gravitational quantum interference has been observed in laboratories, it is essentially to explore more about its theoretical aspects, especially the applications in astronomy. As mentioned above, in TG we can separate the phase $(mc/\hbar)\int \dd s$ into three parts with the third term standing for the gravitational interaction. This term called {\em gravitational phase} is exactly given by 
\be 
\frac{mc}{\hbar}\int u_a {B^a}_\mu\dd x^\mu, 
\label{GP}
\ee
where ${B^a}_\mu$ is a gauge potential associated to gravity~\cite{Aldrovandi:2013wha}. As Aldrovandi {\em et al.} showed~\cite{Aldrovandi:2003pd}, 
in the weak field limit this term gives the same result as the one in the experiment~\cite{Colella:1975dq} for the interference of neutrons on the earth. This coincidence inspired us to apply this expression and its generalization to other scenes, especially the gravitational quantum interference in the Kerr spacetime, to give a prediction for future  observations.

The structure of this paper is arranged as it follows. In Sec.~\ref{TG}, we make a brief introduction to the concept of tetrad in TG. In Sec.~\ref{GPF}, we present a method to calculate the gravitational phase. Its integral expression is derived in the inertial frames and applied to the Kerr spacetime. Therefore, we use this expression for an interference experiment in Sec.~\ref{interference}. Finally in Sec.~\ref{Conclusion}, we summarize the results and present the potential extensions. 

Throughout this article, we use the units $c=G=1$ and the metric signature $(+,-,-,-)$, unless we explicitly specify.

\section{A brief introduction to Teleparallel Gravity\label{TG}}

\subsection{Tetrad in Teleparallel Gravity}
All the formulas in this section are taken from the book~\cite{Aldrovandi:2013wha}, which gives a full introduction to TG. We will not show all the details of this theory, but only introduce the core concepts relevant to our study. Let us start with the tetrad, namely 
\be
h_a={h_a}^\mu\partial_\mu,
\qquad
h^a={h^a}_\mu \dd x^\mu,
\label{tetrad basis}
\ee
a basis which connects the spacetime metric $g_{\mu\nu}$ to the Minkowski's metric in the tangent space,
\be
\eta_{ab}=\text{diag}(1,-1,-1,-1). 
\ee
At each point:
\be
g_{\mu\nu}=\eta_{ab} {h^a}_\mu {h^b}_\nu,
\qquad
\eta_{ab}=g_{\mu\nu}{h_a}^\mu {h_b}^\nu,
\label{metric transformation}
\ee
where the Greek letters are used to denote the coordinates in spacetime, while the Latin letters denote the coordinates in the tangent-space. The components of the tetrad satisfy the equations:
\be
{h^a}_\mu {h_a}^\nu=\delta^\nu_\mu,
\qquad
{h^a}_\mu {h_b}^\mu=\delta^a_b.
\label{tetrad components}
\ee 
Finally, the tetrad relates the spacetime tensors with the tangent-space tensors:
\be
V^\mu={h_a}^\mu V^a,
\qquad
V_a={h^a}_\mu V^\mu.
\label{tensor transformation}
\ee
The components of the tetrad in the presence of gravity are given by:
\be
{h^a}_\mu=\partial_\mu x^a+\dot{A}{^a}_{b\mu}x^b +{B^a}_\mu,
\label{tetrad}
\ee
where $\dot{A}{^a}_{b\mu}={\Lambda^a}_d(x) \partial_\mu {\Lambda_b}^d(x)$ is the Lorentz connection with ${\Lambda^a}_d(x)$ a local Lorentz transformation from an inertial reference frame to a general frame, and ${B^a}_\mu$ is a gauge potential corresponding to a translational transformation $\delta x^a(x)=\varepsilon^a(x)$ on the tangent space. In TG, gravity is generated from the group of the latter transformations under which the tetrad ${h^a}_\mu$ is invariant, while the potential ${B^a}_\mu$ transforms according to
\be
\delta B{^a}_\mu
=-\partial_\mu \varepsilon^a
-\dot{A}{^a}_{b\mu}\varepsilon^b.
\label{gauge B}
\ee
In Eq.~\eqref{tetrad} we see that the expression of the tetrad contains three terms. The first one corresponds to a coordinates' transformation from the spacetime to its tangent-space. As shown in \cite{Aldrovandi:2013wha}, the second one corresponds to the inertia. And the last one corresponds to the gravitational interaction. The expression of the tetrad is obtained by combining \eqref{tetrad} with \eqref{tetrad basis}, namely 
\be
{h^a}=\dd x^a +\dot{A}{^a}_{b\mu}x^b \dd x^\mu +B{^a}_\mu\dd x^\mu.
\label{ha expression}
\ee
Opposite to general relativity, in TG, the curvature vanishes while the torsion is non-vanishing, namely 
\bea
\dot{R}^a{_{b\mu\nu}} &=&\partial_\mu \dot{A}^a{_{b\nu}} -\partial_\nu \dot{A}^a{_{b\mu}} +\dot{A}^a{_{c\mu}} \dot{A}^c{_{b\nu}} -\dot{A}^a{_{c\nu}} \dot{A}^c{_{b\mu}} =0,
\label{dotR}
\\
\dot{T}^a{_{\mu\nu}} &=&\partial_\mu h^a{_\nu} -\partial_\nu h^a{_\mu} +\dot{A}^a{_{c\mu}} h^c{_\nu} -\dot{A}^a{_{c\nu}} h^c{_\mu} 
=\dot{\mathscr{D}}_\mu {B^a}_\nu - \dot{\mathscr{D}}_\nu {B^a}_\mu
\ne 0,
\label{dotT}
\eea
where the derivative operator $\dot{\mathscr{D}}_\mu$ only acts on the indices in the tangent space and it is defined by:
\be
\dot{\mathscr{D}}_\mu \phi^a =\partial_\mu \phi^a +\dot{A}^a{_{b\mu}} \phi^b.
\label{W connection}
\ee
The curvature and the torsion can also be expressed in terms of spacetime indices, i.e. 
\bea
\dot{R}^\rho{_{\lambda\nu\mu}} &=&\partial_\nu \dot{\Gamma}^\rho{_{\lambda\mu}} -\partial_\mu \dot{\Gamma}^\rho{_{\lambda\nu}} +\dot{\Gamma}^\rho{_{\eta\nu}} \dot{\Gamma}^\eta{_{\lambda\mu}} -\dot{\Gamma}^\rho{_{\eta\mu}} \dot{\Gamma}^\eta{_{\lambda\nu}},
\label{dotRspacetime}
\\
\dot{T}^\rho{_{\nu\mu}} &=&\dot{\Gamma}^\rho{_{\mu\nu}} -\dot{\Gamma}^\rho{_{\nu\mu}},
\label{dotTspacetime}
\eea
where $\dot{\Gamma}^\mu{_{\rho\nu}}$ is the Weitzenb\"ock connection defined by:
\be
\dot{\Gamma}^\mu{_{\rho\nu}}= {h_a}^\mu \dot{\mathscr{D}}_\nu {h^a}_\rho.
\label{GammaDefine}
\ee  

In TG, the torsion is regarded as a field strength, and from \eqref{dotT} we see that $B^a{_\mu}$ plays a role analogous to the gauge potential in electromagnetism. The torsion is gauge invariant~\cite{Aldrovandi:2013wha} because it can be written in the following form, 
\be
\dot{T}^a{_{\mu\nu}}=\dot{\mathscr{D}}_\mu {h^a}_\nu - \dot{\mathscr{D}}_\nu {h^a}_\mu ,
\ee
while the tetrad is invariant under the gauge transformation \eqref{gauge B}. The action for gravity is constructed by means of the torsion tensor, which coincides with the Einstein-Hilbert action, and the field equation in TG is equivalent to the Einstein equation (all the details can be found in the book~\cite{Aldrovandi:2013wha}).

\subsection{The role of the gauge potential\label{Role of B}}
As mentioned above, the gravitational phase is given by \eqref{GP} where the gauge potential $B^a{_\mu}$ appears in the integrand. This is reasonable because gravity is represented by the gauge potential, as stated in Ref.~\cite{Aldrovandi:2013wha}. This potential not only appears in the field equation, but also plays an important role in the equation of motion, which is equivalent to the geodesic equation, of a particle in the gravitational field. We now prove the latter claim and finally show that $B^a{_\mu}$ appears in the gravitational phase by an analogy with electromagnetism.

Let us remind the geodesic equation in general relativity, namely 
\be
\frac{\dd u^\mu}{\dd s} + \Gamma^\mu{_{\rho\nu}} u^\rho u^\nu =0.
\label{geodesic}
\ee
In TG, the Levi-Civita connection can be written as~\cite{Aldrovandi:2013wha}
\be
\Gamma^\mu{_{\rho\nu}}
=\dot{\Gamma}^\mu{_{\rho\nu}}-\dot{K}^\mu{_{\rho\nu}},
\label{GammaK}
\ee
where $\dot{\Gamma}^\mu{_{\rho\nu}}$ is defined in \eqref{GammaDefine}, and $\dot{K}^\mu{_{\rho\nu}}$ is the contortion
\be
\dot{K}^\mu{_{\rho\nu}}
=\frac{1}{2} (\dot{T}_\nu{^\mu}_\rho +\dot{T}_\rho{^\mu}_\nu 
-\dot{T}^\mu{_{\rho\nu}}  ),
\label{Kdefiniton}
\ee
of the Weitzenb\"ock torsion
\be
\dot{T}^\mu{_{\rho\nu}}={h_a}^\mu \dot{T}^a{_{\rho\nu}}
={h_a}^\mu (\dot{\mathscr{D}}_\rho {B^a}_\nu - \dot{\mathscr{D}}_\nu {B^a}_\rho),
\label{Tdefine}
\ee
where \eqref{dotT} has been used. Recalling \eqref{tetrad}, the tetrad depends on ${B^a}_\mu$. Therefore, both $\dot{\Gamma}^\mu{_{\rho\nu}}$ in \eqref{GammaDefine} and $\dot{K}^\mu{_{\rho\nu}}$ in \eqref{Kdefiniton} depend on the gauge potential. 

According to the above expressions, we can prove that the geodesic equation \eqref{geodesic} depends on the potential ${B^a}_\mu$. Indeed, we can rewrite the geodesic equation in TG using \eqref{GammaK}, 
\be
\frac{\dd u^\mu}{\dd s} + (\dot{\Gamma}^\mu{_{\rho\nu}} -\dot{K}^\mu{_{\rho\nu}} ) u^\rho u^\nu =0.
\label{EOMu}
\ee
For the last term, according to \eqref{Kdefiniton}, we get
\be
\dot{K}^\mu{_{\rho\nu}} u^\rho u^\nu
=\frac{1}{2} (\dot{T}_\nu{^\mu}_\rho +\dot{T}_\rho{^\mu}_\nu 
-\dot{T}^\mu{_{\rho\nu}}  ) u^\rho u^\nu
=\dot{T}_\rho{^\mu}_\nu u^\rho u^\nu,
\label{KequalT}
\ee
where the last step follows from the anti-symmetry of $\dot{T}^\mu{_{\rho\nu}}$ in the last two indices (see \eqref{Tdefine}), and by re-labeling the indices of the first term. Furthermore, we rewrite \eqref{KequalT} as:
\bea
\dot{K}^\mu{_{\rho\nu}} u^\rho u^\nu
&=&g_{\rho\alpha} g^{\mu\beta} \dot{T}^\alpha{_{\beta\nu}} u^\rho u^\nu
\nonumber\\
&=&(\eta_{cd} {h^c}_\rho {h^d}_\alpha) (\eta^{ef} {h_e}^\mu  {h_f}^\beta ) 
{h_a}^\alpha (\dot{\mathscr{D}}_\beta {B^a}_\nu - \dot{\mathscr{D}}_\nu {B^a}_\beta) u^\rho u^\nu
\nonumber\\
&=&{h_a}^\mu
(\eta_{ce} {h^c}_\rho ) (\eta^{af} {h_f}^\beta )
 (\dot{\mathscr{D}}_\beta {B^e}_\nu - \dot{\mathscr{D}}_\nu {B^e}_\beta) u^\rho u^\nu,
 \label{Kuu}
\eea
where \eqref{metric transformation} and \eqref{Tdefine} have been used in the second step and \eqref{tetrad components} has been used in the last step. Then plugging \eqref{GammaDefine} and \eqref{Kuu} into \eqref{EOMu}, we get the equation of motion for a point-like particle:
\be
\frac{\dd u^\mu}{\dd s} 
+ {h_a}^\mu\Bigl[ \dot{\mathscr{D}}_\nu {h^a}_\rho
-(\eta_{ce} {h^c}_\rho ) (\eta^{af} {h_f}^\beta )
 (\dot{\mathscr{D}}_\beta {B^e}_\nu - \dot{\mathscr{D}}_\nu {B^e}_\beta)
\Bigr] u^\rho u^\nu =0,
\label{duds}
\ee
which is equivalent to the geodesic equation \eqref{geodesic}. 

We now show by contradiction that in presence of gravity the gauge potential can not be eliminated from the equation \eqref{duds}. 
We first replace  \eqref{tetrad} in \eqref{duds} and afterwards assume the gauge potential to vanish. Hence, we rewrite \eqref{duds} in cartesian coordinates of an inertial frame in which the Lorentz connection $\dot{A}{^a}_{b\mu}$ vanishes and the tetrad components take the form ${h^a}_\rho={\delta^a}_\rho$ (see Ref.~\cite{Aldrovandi:2013wha}). Therefore, the equation \eqref{duds} simplifies to:
\be
\frac{\dd u^\mu}{\dd s} 
+ {h_a}^\mu (\partial_\nu {\delta^a}_\rho) u^\rho u^\nu =0
\qquad
\Longrightarrow \quad
\frac{\dd u^\mu}{\dd s} =0,
\label{duds0}
\ee
where we used the definition \eqref{W connection} and $\partial_\nu {\delta^a}_\rho=0$. Therefore, in cartesian coordinates of an inertial frame and assuming that \eqref{duds} does not depend on the gauge potential, equation \eqref{duds} reduces to the equation of a free particle. On the other hand, we know that in the presence of gravity \eqref{duds} does not reduce to the equation of a free particle because it is equivalent to the geodesic equation (\ref{geodesic}). Therefore, in the presence of gravity we can not eliminate the gauge potential from the equation \eqref{duds} and the gauge potential ${B^a}_\mu$ represents the effect of gravity on the motion of a point-like particle. 

We would also emphasize the role of the Lorentz connection $\dot{A}^a{_{b\mu}}$. As stated in Ref.~\cite{Aldrovandi:2013wha}, this connection is due to the inertial effects and it appears in the tetrad when a general reference is chosen. Hence, in this case, it also appears in the equation of motion. However, if we take an inertial frame, this connection vanishes. Indeed, such connection is constructed with the local Lorentz transformation ${\Lambda^a}_d(x)$ from an inertial frame to a general frame, namely $\dot{A}{^a}_{b\mu}={\Lambda^a}_d(x) \partial_\mu {\Lambda_b}^d(x)$. In particular, since the Lorentz transformation from an inertial frame to another inertial frame is a global transformation, $\dot{A}{^a}_{b\mu}$ vanishes in the inertial frames. 

Therefore, based on the above discussions, generally, the motion of the particle is governed by both the Lorentz connection 
$\dot{A}^a{_{b\mu}}$ and the gauge potential ${B^a}_\mu$. If an inertial frame is chosen, the motion is only governed by the later. These two quantities together plays a role similar to the Levi-Civita connection in general relativity. Indeed, in general relativity, the motion of the particle is governed by the Levi-Civita connection, as the equation \eqref{geodesic} shows.

Finally, let us show that the gauge potential $B^a{_\mu}$ appears in the gravitational phase, though we have proved that it affects the equation of motion of the particle.  As shown in the Ref.~\cite{Aldrovandi:2013wha}, the equation of motion~\eqref{EOMu} can be derived directly from the following action principle, 
\be
\mathscr{S}=-m\int_p^q  (u_a \dd x^a+u_a \dot{A}{^a}_{b\mu}x^b \dd x^\mu +u_a B{^a}_\mu\dd x^\mu),
\label{S action}
\ee
where the first term stands for the free particle, the second term relates to the inertial effects, and the last term represents the gravitational interaction. Here $u_a=\eta_{ab}u^b$ and $u^b$ is a four-velocity defined in the tangent space (see \eqref{four velocities}). In presence of the  electromagnetic potential $A_\mu$, the action \eqref{S action}, for a charged particle of charge $q$, should be modified by adding the term $(q/m) A_\mu \dd x^\mu$ under the integral in (\ref{S action})~\cite{Aldrovandi:2013wha}. In special relativity, the action of a particle in presence of the electromagnetic field is just the combination of a free term and the interaction term with the electromagnetic potential. Correspondingly, the electromagnetic phase factor for an Aharonov-Bohm effect~\cite{Aharonov:1959fk} is given by $e^{\frac{i}{\hbar} \int q A_\mu dx^\mu}$. Thus, for a gravitational field, in strict analogy with the electromagnetism, the last two terms in \eqref{S action} contribute to the gravitational phase factor~\cite{Aldrovandi:2013wha}. Especially, if we choose an inertial frame, the second term in \eqref{S action} vanishes and only the last term contributes to the gravitational phase factor. In this frame, the gauge potential $B^a{_\mu}$ dominates the gravitational phase. Indeed, as we see from the definition of the field strength~\eqref{dotT}, the role of the potential $B^a{_\mu}$ in gravity is similar to the role of the gauge potential in electromagnetism. It deserves to be mentioned that a similar discussion of the gravitational phase can be found in Ref.~\cite{Aldrovandi:2003pd}.

\section{Gravitational phase\label{GPF}}
In this section we first provide the general formula for the gravitational phase and afterwards we evaluate it explicitly for the case of the Kerr spacetime. 

\subsection{Gravitational phase in inertial references}

The gravitational phase factor for a massive particle in a generic frame is~\cite{Aldrovandi:2013wha,Aldrovandi:2003pd}:
\be
\Phi_g =\exp \Bigl(-\frac{i}{\hbar} \mathscr{S}_g \Bigr),
\label{Phi g}
\ee
where
\be
\mathscr{S}_g=-m\int_p^q u_a (\dot{A}{^a}_{b\mu}x^b \dd x^\mu +B{^a}_\mu\dd x^\mu)
\label{Sg}
\ee
is the interaction part of the action
\be
\mathscr{S}=-m\int_p^q \dd s,
\label{action int}
\ee
and the four-velocities in spacetime and tangent-space are defined respectively as follows, 
\be
u^\mu=\frac{\dd x^\mu}{\dd s},
\qquad
u^a=\frac{h^a}{\dd s}.
\label{four velocities}
\ee

For simplicity, we choose an inertial coordinate system $K$ in which $\dot{A}{^a}_{b\mu}=0$. Hence, according to \eqref{Sg}, the interaction action reads:
\be
\mathscr{S}_g=-m\int_p^q u_a B{^a}_\beta \dd x^\beta
=-m\int_p^q g_{\mu\nu}u^\mu B{^\nu}_\beta \dd x^\beta,
\label{SgKK}
\ee
where the second equation in \eqref{tensor transformation} is used and the function $B{^\nu}_\beta$ is defined as
\be
B{^\nu}_\beta=h_a{^\nu}B{^a}_\beta
=(h^T B){^\nu}_\beta.
\label{B spacetime}
\ee
Here $h^T$ is the transpose matrix of $h_a{^\nu}$, and $B$ is the matrix $B{^a}_\beta$.
Therefore, if we have the expressions for $g_{\mu\nu}$, $u^\mu$ and ${B^\nu}_\beta$, we can evaluate $\mathscr{S}_g$. Plugging \eqref{SgKK} into \eqref{Phi g}, we get the gravitational phase factor for a massive particle:
\be
\Phi_g=\exp\Bigl(\frac{i}{\hbar}m\int_p^q g_{\mu\nu}u^\mu B{^\nu}_\beta \dd x^\beta\Bigr).
\label{Phi mass}
\ee

For massless particles, let us consider the light firstly. For a light, its phase factor can be written as:
\be
\Phi=\exp (i\psi)
=\exp\Bigl(\frac{i}{\hbar}\int_p^q P_\mu \dd x^\mu \Bigr),
\label{Phi L}
\ee
where $P_\mu=\hbar k_\mu$ is the four-momentum of the photon, and $k_\mu$ is the wave vector. In Ref.~\cite{Stodolsky:1978ks}, the optical interferometry is based on \eqref{Phi L}, but for a weak gravitational field. Unlike in the Ref.~\cite{Stodolsky:1978ks}, we extract the gravitational part from the phase factor in the framework of TG, without need of the weak field approximation. According to \eqref{Phi L}, we have:
\be
\dd \psi
=k_\mu\dd x^\mu
=k_a h^a
=k_a (\dd x^a +\dot{A}{^a}_{b\mu}x^b \dd x^\mu +B{^a}_\mu\dd x^\mu) \, ,
\label{d psi}
\ee
where Eqs.~\eqref{tensor transformation}, \eqref{metric transformation}, \eqref{tetrad components}, and \eqref{ha expression} have been used. Since we only need the interaction part, for the gravitational phase $\phi_g$ we have: 
\be
\dd \phi_g
=k_a (\dot{A}{^a}_{b\mu}x^b \dd x^\mu +B{^a}_\mu\dd x^\mu).
\ee
Moreover, if we choose an inertial frame for which $\dot{A}{^a}_{b\mu}=0$, the gravitational phase simplifies to: 
\be
\phi_g=\int_p^q k_a B{^a}_\mu\dd x^\mu
=\int_p^q k_\nu {B^\nu}_\mu\dd x^\mu,
\label{psi g}
\ee
where \eqref{tensor transformation} is used and $B{^\nu}_\beta$ is defined in \eqref{B spacetime}. 
Finally, the gravitational phase factor for light is:
\be
\Phi_L
=\exp(i \phi_g)
=\exp\Bigl(\frac{i}{\hbar} \int_p^q 
g_{\mu\nu}P^\mu {B^\nu}_\beta\dd x^\beta
\Bigr).
\label{Phi light}
\ee
Although \eqref{Phi light} has been derived for photons, we assume it also applicable for other massless particles. Of course, this hypothesis needs a rigorous proof.

In summary, the gravitational phase for a particle (massive or massless) in an inertial frame is given by:
\be
\phi_g=\frac{1}{\hbar}\int_p^q S_\beta \dd x^\beta,
\label{phi_g}
\ee
where the function $S_\beta$ is defined as
\be
S_\beta=g_{\mu\nu}P^\mu {B^\nu}_\beta,
\label{defining S}
\ee
and $P^\mu$ is the four-momentum. The gravitational phase factor is given by $\Phi_g=\exp(i\phi_g)$.

To calculate $S_\beta$, we need to know ${B^\nu}_\beta$ firstly. According to \eqref{B spacetime}, the expression of ${B^\nu}_\beta$ is given by ${h_a}^\nu$ and $B{^a}_\beta$. Thus in addition to ${h_a}^\nu$, we need to seek the expression for $B{^a}_\beta$. Before proceeding, let us consider the cartesian coordinate system $K'$ in in which $\partial_{\mu'} x^a=\delta{^a}_{\mu'}$ holds~\cite{Aldrovandi:2013wha}. Therefore, according to Eq.~\eqref{tetrad}, in the coordinate $K'$ the gauge potential can be written as:
 \be
 B{^a}_{\mu'}=h{^a}_{\mu'}-\delta{^a}_{\mu'}.
 \label{fieldK}
 \ee
Moreover, the components of the tetrad in the generic coordinate~$K$ can be expressed as~\cite{Pereira:2001xf}:
\be
h{^a}_\rho =h{^a}_{\nu'} \frac{\partial x^{\nu'}}{\partial x^\rho} \, ,
\label{hhprime}
\ee
which can be derived directly by writing the second equation of \eqref{tetrad basis} as:
\bea
h^a= h{^a}_{\nu'} \dd x^{\nu'}
=h{^a}_{\nu'} \frac{\partial x^{\nu'}}{\partial x^\rho}\dd x^\rho
=h{^a}_\rho \dd x^\rho \, .
\eea
Now we come back to the expression for $B{^a}_{\mu}$.  We write the gravitational phase in the coordinate~$K$:
\be
\phi_g
=\frac{1}{\hbar}\int_p^q g_{\mu\nu}P^\mu {B^\nu}_\beta \dd x^\beta
=\frac{1}{\hbar}\int_p^q P_a B{^a}_{\beta} \dd x^{\beta},
\label{SgK0}
\ee
where \eqref{tensor transformation} and \eqref{tetrad components} are used. On the other hand, we write it in the cartesian coordinate~$K'$:
\be
\phi_g
=\frac{1}{\hbar}\int_p^q P_a B{^a}_{\mu'} \dd x^{\mu'}
=\frac{1}{\hbar}\int_p^q P_a (h{^a}_\sigma\frac{\partial x^\sigma}{\partial x^{\mu'}}-\delta{^a}_{\mu'}) \dd x^{\mu'},
\label{SgKprime}
\ee
where \eqref{fieldK} and \eqref{hhprime} are used. Furthermore, we write \eqref{SgKprime} as:
\be
\phi_g
=\frac{1}{\hbar}\int_p^q P_a (h{^a}_\sigma\frac{\partial x^\sigma}{\partial x^{\mu'}}-\delta{^a}_{\mu'}) \frac{\partial x^{\mu'}}{\partial x^\beta}\dd x^\beta
=\frac{1}{\hbar}\int_p^q P_a (h{^a}_\beta-\delta{^a}_{\mu'} \frac{\partial x^{\mu'}}{\partial x^\beta})\dd x^\beta.
\label{SgK}
\ee
Comparing \eqref{SgK0} and \eqref{SgK}, we finally obtain:
\be
B{^a}_\beta=h{^a}_\beta -\delta{^a}_{\mu'}\frac{\partial x^{\mu'}}{\partial x^\beta}.
\label{B field}
\ee

Summarizing. We choose an inertial coordinate system $K$. Then we find the expression for the components of the tetrad ${h^a}_\beta$, and the transformation between the coordinate $K$ and the cartesian coordinate $K'$. Plugging the tetrad into \eqref{B field}, we get the expression of ${B^a}_\beta$. Hence, inserting the latter into \eqref{B spacetime}, we get ${B^\nu}_\beta$. Pugging the expressions of ${B^\nu}_\beta$ and $P^\mu$ into \eqref{defining S}, we get $S_\beta$. According to it, we calculate the integral in \eqref{phi_g} to finally get the gravitational phase.

\subsection{Gravitational phase in the Kerr spacetime}
Using Boyer-Lindquist coordinates $K (t, r, \theta, \varphi)$ in the Kerr spacetime, the matrix form of the metric is~\cite{Chandrasekhar1998}:
\be
(g_{\mu\nu})=
\begin{pmatrix}
g_{00}&{}&{}&g_{03}\\
{}&g_{11}&{}&{}\\
{}&{}&g_{22}&{}\\
g_{30}&{}&{}&g_{33}
\end{pmatrix}
=
\begin{pmatrix}
1-r_g r/\rho^2&{0}&{0}&a r_g r(\sn\theta)^2/\rho^2\\
{0}&-\rho^2/\Delta&{0}&{0}\\
{0}&{0}&-\rho^2&{0}\\
a r_g r(\sn\theta)^2/\rho^2&{0}&{0}&-[r^2+a^2+a^2 r_g r(\sn\theta)^2/\rho^2](\sn\theta)^2
\end{pmatrix},
\label{metric}
\ee
and for the inverse:
\be
(g^{\mu\nu})=
\begin{pmatrix}
g^{00}&{}&{}&g^{03}\\
{}&g^{11}&{}&{}\\
{}&{}&g^{22}&{}\\
g^{30}&{}&{}&g^{33}
\end{pmatrix}
=
\begin{pmatrix}
\Sigma^2/(\rho^2\Delta)&{0}&{0}&a r_g r/(\rho^2\Delta)\\
{0}&-\Delta/\rho^2&{0}&{0}\\
{0}&{0}&-1/\rho^2&{0}\\
a r_g r/(\rho^2\Delta)&0&0&-[\Delta-a^2(\sn\theta)^2]/[\rho^2(\sn\theta)^2\Delta]
\end{pmatrix},
\label{inverse metric}
\ee
where 
\be
r_g=2M, \quad
\rho^2=r^2+a^2(\cn\theta)^2, \quad
\Delta=r^2-r_g r +a^2, \quad
\Sigma^2=(r^2+a^2)^2-a^2(\sn \theta)^2\Delta . 
\label{rgrho}
\ee
The parameter $M$ is the mass of the black hole, and $a$ is its angular momentum per unit of mass in the units $c=1$. (In the SI units it is $a=J/(Mc)$, where $J$ is the angular momentum of the black hole~\cite{Landau:1975pou}.) Here the symbols $\sn\theta$, $\cn\theta$, $\sn\varphi$ and $\cn\varphi$ denote $\sin(\theta)$, $\cos(\theta)$, $\sin(\varphi)$ and $\cos(\varphi)$ respectively.

We will calculate the gravitational phase in the Kerr spacetime by using the last expression in \eqref{SgKK}, but before that, we derive the expression of $B{^a}_\beta$ according to Eq.~\eqref{B field}. 
The coordinate transformation from $K'$ to $K$ is~\cite{Landau:1975pou}:
\be
\begin{cases}
t'=t,\\
x'=\sqrt{r^2+a^2}\, \sn\theta\cn\varphi,\\
y'=\sqrt{r^2+a^2}\, \sn\theta\sn\varphi,\\
z'=r\, \cn\theta.
\end{cases}
\label{transformation}
\ee
From Eq.~\eqref{transformation} we can get the Jacobi matrix:
\be
\Bigl(\frac{\partial x^{\mu'}}{\partial x^\beta}\Bigr)=
\begin{pmatrix}
1 & 0 & 0 &0\\
0& \frac{r}{\rho_0}\sn\theta \cn\varphi &\rho_0 \cn\theta\cn \varphi &-\rho_0\sn\theta\sn\varphi\\
0& \frac{r}{\rho_0}\sn\theta \sn\varphi &\rho_0 \cn\theta\sn \varphi &\rho_0\sn\theta\cn\varphi\\
0&\cn\theta &-r\sn\theta &0
\end{pmatrix},
\label{Jacobi}
\ee
where $\rho_0=\sqrt{r^2+a^2}$. The tetrad in the Kerr spacetime is~\cite{Aldrovandi:2013wha,Pereira:2001xf}:
\be
(h{^a}_\beta)=
\begin{pmatrix}
\gamma_{00} & 0 & 0 &\eta\\
0& \gamma_{11}\sn\theta \cn\varphi & \gamma_{22}\cn\theta\cn\varphi &-\zeta\sn\varphi\\
0&  \gamma_{11}\sn\theta\sn\varphi & \gamma_{22}\cn\theta\sn\varphi &\zeta\cn\varphi\\
0& \gamma_{11}\cn\theta & -\gamma_{22}\sn\theta &0
\end{pmatrix},
\label{tetrad Kerr}
\ee
where\footnote{In Ref.~\cite{Pereira:2001xf} the authors only give the expression $\zeta^2=\eta^2-g_{33}$. We believe $\zeta=\sqrt{\eta^2-g_{33}}$ also holds, which is in accordance with the tetrad in Schwardschild space time (see (29) in Ref.~\cite{Pereira:2001xf}).}
\be
\eta=g_{03}/\gamma_{00},\quad
\zeta=\sqrt{\eta^2-g_{33}},\quad
\gamma_{00}=\sqrt{g_{00}},\quad
\gamma_{jj}=\sqrt{-g_{jj}}.
\label{some quantities}
\ee 
Inserting Eqs. \eqref{tetrad Kerr} and \eqref{Jacobi} into \eqref{B field}, we get the gauge potential in the Kerr spacetime:
\be
(B{^a}_\beta)=
\begin{pmatrix}
\gamma_{00}-1 & 0 & 0 &\eta\\
0& (\gamma_{11}-\frac{r}{\rho_0})\sn\theta\cn\varphi & (\gamma_{22}-\rho_0)\cn\theta\cn\varphi &(\rho_0\sn\theta-\zeta)\sn\varphi\\
0&  (\gamma_{11}-\frac{r}{\rho_0})\sn\theta\sn\varphi & (\gamma_{22}-\rho_0)\cn\theta\sn\varphi &(\zeta-\rho_0\sn\theta)\cn\varphi\\
0& (\gamma_{11}-1)\cn\theta & (r-\gamma_{22})\sn\theta &0
\end{pmatrix}.
\label{B Kerr}
\ee
It is easy to check that ${B^a}_\beta=0$ in flat spacetime, namely for $a=0$ and $M=0$. The matrix form for the inverse of the tetrad is\footnote{We think that these are some typos in equation (14.37) in~\cite{Aldrovandi:2013wha}. This equation should be modified as \eqref{inverse tetrad}.}:
\be
(h_a{^\nu})=
\begin{pmatrix}
\gamma_{00}^{-1} & 0 & 0 &0\\
\zeta g^{03}\sn\varphi & \gamma_{11}^{-1}\sn\theta \cn\varphi & \gamma_{22}^{-1}\cn\theta\cn\varphi &-\zeta^{-1}\sn\varphi\\
-\zeta g^{03}\cn\varphi &  \gamma_{11}^{-1}\sn\theta\sn\varphi & \gamma_{22}^{-1}\cn\theta\sn\varphi &\zeta^{-1}\cn\varphi\\
0& \gamma_{11}^{-1}\cn\theta & -\gamma_{22}^{-1}\sn\theta &0
\end{pmatrix}.
\label{inverse tetrad}
\ee
One can check that \eqref{tetrad Kerr} and \eqref{inverse tetrad} indeed satisfy $g_{\mu \nu }=\eta_{ab} {h^a}_\mu {h^b}_\nu$, $\eta_{ab}=g_{\mu \nu } {h_a}^\mu {h_b}^\nu$ and ${h^a}_\mu {h_b}^{\mu }=\delta^a_b$. Inserting Eqs.~\eqref{B Kerr} and \eqref{inverse tetrad} into \eqref{B spacetime}, we obtain
\be
(B{^\nu}_\beta)=
\begin{pmatrix}
1-\gamma_{00}^{-1} & 0 & 0 &\eta\gamma_{00}^{-1} +(\rho_0\sn\theta-\zeta)\zeta g^{03}\\
0& 1-\gamma_{11}^{-1}[ r \rho_0^{-1}(\sn\theta)^2+(\cn\theta)^2] & \gamma_{11}^{-1}\sn\theta\cn\theta(r-\rho_0) &0\\
0&  \gamma_{22}^{-1}\sn\theta\cn\theta(1-r/\rho_0)& 1-\gamma_{22}^{-1}[r (\sn\theta)^2+\rho_0 (\cn\theta)^2] &0\\
0&0 &0 &1-\zeta^{-1}\rho_0\sn\theta
\end{pmatrix}.
\label{B expression}
\ee

In terms of Eq.~\eqref{defining S}, the first expression in Eq.~\eqref{metric}, and Eq.~\eqref{B expression}, the matrix $S_\beta$ can be written as:
\be
(S_\beta)=(P^\mu g_{\mu\nu}B{^\nu}_\beta)=
\begin{pmatrix}
(1-\gamma_{00}^{-1})(P^0 g_{00}+P^3 g_{30})\\
P^1 g_{11}\{1-\gamma_{11}^{-1}[ r \rho_0^{-1}(\sn\theta)^2+(\cn\theta)^2 ]\} +P^2 g_{22}\gamma_{22}^{-1}\sn\theta\cn\theta(1-r/\rho_0)\\
P^1 g_{11} \gamma_{11}^{-1}\sn\theta\cn\theta(r-\rho_0)  +P^2 g_{22} \{ 1-\gamma_{22}^{-1}[ r(\sn\theta)^2+\rho_0 (\cn\theta)^2]\}\\
 (P^0 g_{00}+P^3 g_{30})[\eta\gamma_{00}^{-1}+(\rho_0\sn\theta-\zeta)\zeta g^{03}]+(P^0 g_{03}+P^3 g_{33})(1-\zeta^{-1}\rho_0\sn\theta)
\end{pmatrix}.
\label{S beta}
\ee
The latter result can be further simplified by using the following conserved quantities in the Kerr spacetime~\cite{Chandrasekhar1998}, 
\bea
E&=&\Bigl(1-\frac{r_g r}{\rho^2}\Bigr)\frac{\dd t}{\dd \xi} +\frac{a r_g r (s\theta)^2}{\rho^2}\frac{\dd \varphi}{\dd \xi}
=u^0 g_{00} +u^3 g_{30}, \nonumber\\
-L&=&\frac{a r_g r (s\theta)^2}{\rho^2}\frac{\dd t}{\dd \xi}-\Bigl[r^2+a^2+\frac{r_g r}{\rho^2}a^2 (s\theta)^2\Bigr] (s\theta)^2 \frac{\dd\varphi}{\dd \xi}
=u^0 g_{03} +u^3 g_{33},
\label{conserved quantities}
\eea
where $\xi$ is an affine parameter (for massive particles it is the proper time), and $E$ and $L$ are defined as
\be
E=
\begin{cases}
\mathcal{E} m^{-1}, &\text{for massive particles,}\\
\mathcal{E}, &\text{for massless particles,}
\end{cases}
\qquad
L=
\begin{cases}
\mathcal{L} m^{-1}, &\text{for massive particles,}\\
\mathcal{L}, &\text{for massless particles.}
\end{cases}
\label{defining EL}
\ee
Notice that for massive particles we have $P^\mu=m u^\mu$, while for massless particles we have $P^\mu=u^\mu$. Here the quantity $\mathcal{E}$ has the meaning of energy, while the quantity $\mathcal{L}$ has the meaning of angular momentum along the spin of the black hole. Plugging \eqref{conserved quantities} into \eqref{S beta}, we get
\be
(S_\beta)=
\begin{pmatrix}
(1-\gamma_{00}^{-1})\mathcal{E}\\
P^1 g_{11}\{ 1-\gamma_{11}^{-1}[ r \rho_0^{-1}(\sn\theta)^2+(\cn\theta)^2] \} +P^2 g_{22}\gamma_{22}^{-1}\sn\theta \cn\theta(1-r/\rho_0)\\
P^1 g_{11} \gamma_{11}^{-1}\sn\theta \cn\theta (r-\rho_0)  +P^2 g_{22} \{ 1-\gamma_{22}^{-1}[r(\sn\theta)^2+\rho_0 (\cn\theta)^2]\}\\
 \mathcal{E}[\eta\gamma_{00}^{-1}+(\rho_0\sn\theta-\zeta)\zeta g^{03}]-\mathcal{L}(1-\zeta^{-1}\rho_0\sn\theta)
\end{pmatrix}.
\label{S beta2}
\ee
Finally, recalling \eqref{phi_g}, the gravitational phase in the Kerr spacetime is given by:
\be
\phi=\frac{1}{\hbar}\int_p^q S_\beta \dd x^\beta.
\label{PhiK}
\ee

\section{Particles interference experiment\label{interference}}
\subsection{Theoretical prediction\label{TPre}}
\begin{figure}[h]
\centering
\includegraphics[scale=0.13]{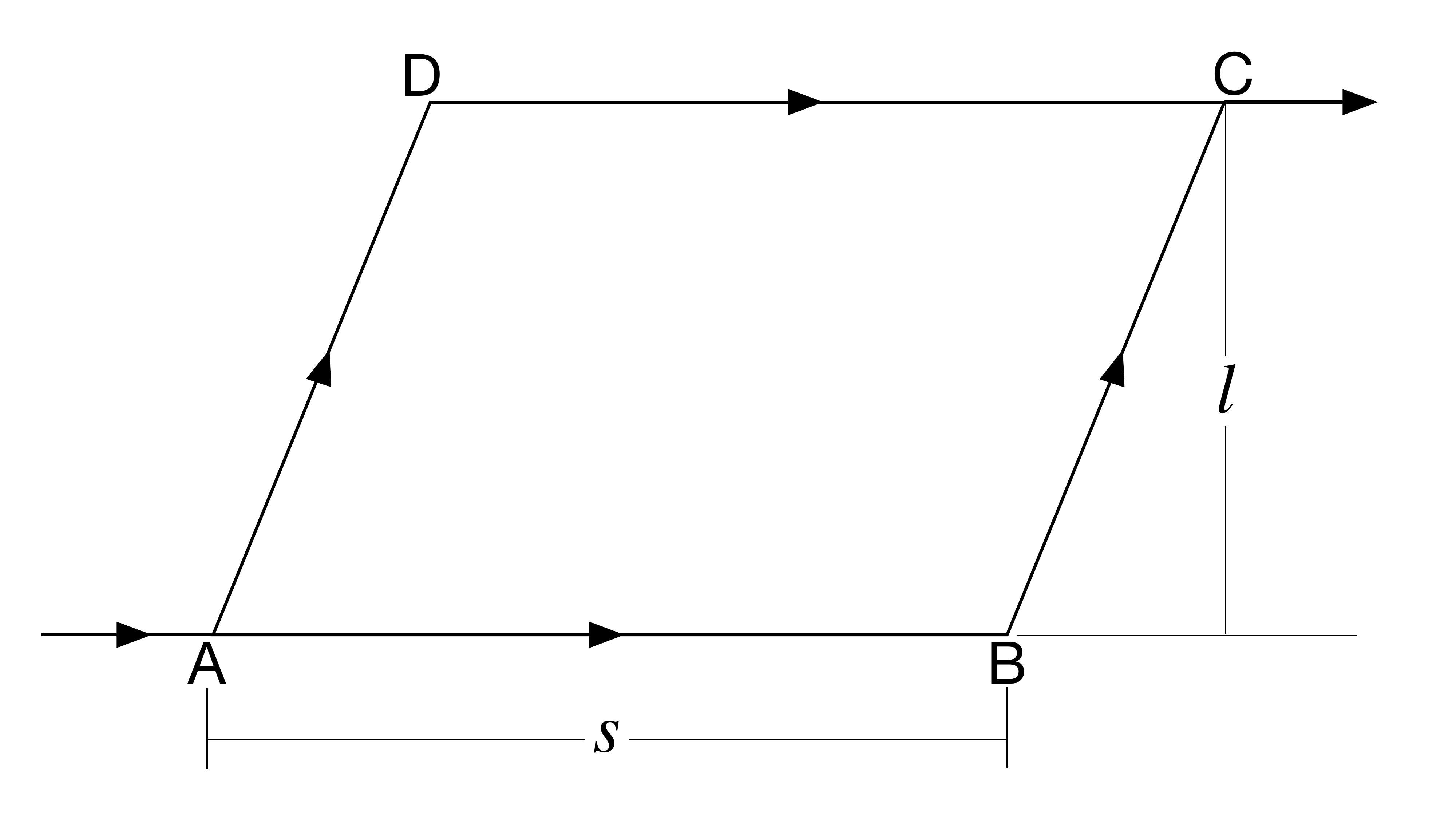}
\caption{Schematic figure for the COW experiment. A beam of neutrons is split into two beams along the sides of a parallelogram which is vertical to the surface of the earth, and, afterwards they interfere.}
\label{COW experiment}
\end{figure}

We will study an interference experiment in the region $r\gg r_g$ with the size of the setup much smaller than its distance from the black hole. Let us start with a review of the Colella-Overhauser-Werner (COW) experiment on the earth~\cite{Colella:1975dq}. The principle of this experiment is shown in FIG.~\ref{COW experiment}, where the parallelogram is vertical and its base AB is parallel to the surface of the earth. A beam of neutrons is split into two beams along the paths ABC and ADC respectively, and, afterwards they interfere. 
Since of the presence of gravity, the phase accumulated along the path ABC is different from the phase accumulated along ADC. The theoretical predictions for this experiment were given in Ref.~\cite{Overhauser:1974}, in which the gravitational phase difference between these two paths was found to be:
\be
\delta\phi=\frac{m^2 g l \lambda_d}{2\pi\hbar^2}s,
\label{COWpre}
\ee
where $m$ is the mass of the neutron, $\lambda_d=2\pi\hbar/(mv)$ is its de Broglie wavelength, $g$ is the gravitational acceleration, $l$ is the height of the parallelogram, and $s$ is the length of AB.  

\begin{figure}[h]
\centering
\includegraphics[scale=0.15]{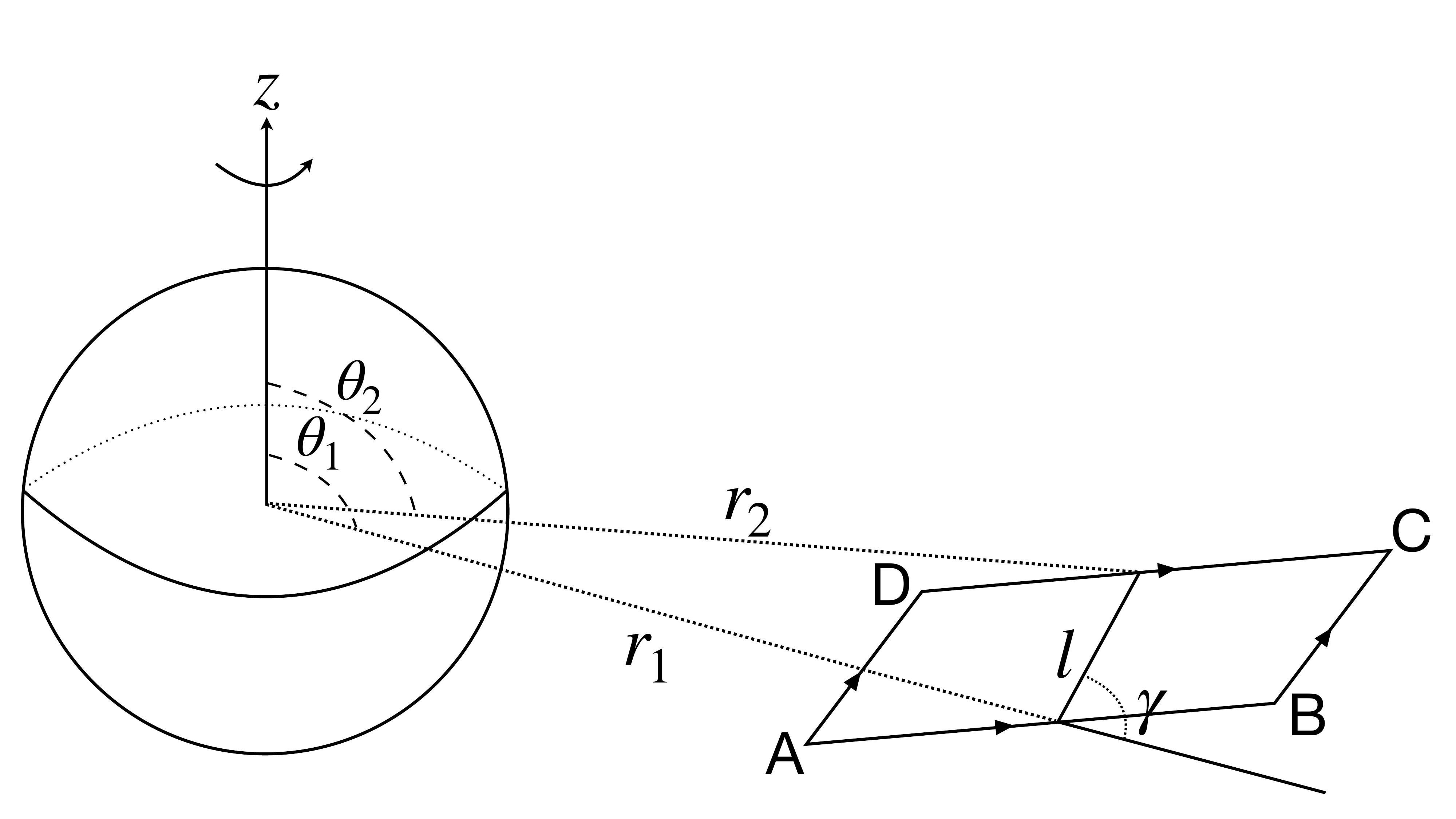}
\caption{The interference experiment in the region $r\gg r_g$ of the Kerr spacetime. The particles are split into two beams along the paths ABC and ADC respectively to afterward interfere. In $r\gg r_g$, we can use the Schwarzschild coordinates to approximate the Boyer-Lindquist coordinates. The axis $z$ of the black hole and the vectors $\vec{r}_1$ and $\vec{r}_2$ are in the same plane, which is perpendicular to the base AB. The angle between $\vec{r}_1$ and the plane of the parallelogram ABCD is $\gamma$. The length of AB is $s$, and the height of the parallelogram is $l$. }
\label{experimentKerr}
\end{figure}

Let us now to place the parallelogram in the region of the Kerr spacetime for $r\gg r_g$. The particles are not limited to be neutrons and the devise is shown in Fig~\ref{experimentKerr}. For simplicity, we assume:\\
(a) The size of the parallelogram to be much smaller than its distance from the black hole, so that the coordinates $r$ and $\theta$ are approximately constant along the paths AB and DC;\\
(b) The energy $\mathcal{E}$ of the particle is conserved even when the particle turns direction at the points B and D (The quantity $\mathcal{L}$ changes at these points, but it is conserved on the paths AB, BC, AD and DC), such that the magnitude of its velocity (defined in \eqref{vsquare}) does not change at such points.

Combining the assumption (a) with \eqref{PhiK}, we write the accumulated gravitational phase along the path AB as 
\bea
\phi_{AB} &\approx&
\frac{1}{\hbar}\Bigl(
\int S_0 \dd t +\int S_3 \dd \varphi 
\Bigr)
\nonumber\\
&=&\frac{1}{\hbar}
(S_0 t_{AB} + S_3 \varphi_{AB})
\nonumber\\
&=&
\frac{1}{\hbar} \mathcal{E} (1-\gamma_{00}^{-1}) t_{AB}
+\frac{1}{\hbar} \Bigl\{ 
\mathcal{E}[\eta\gamma_{00}^{-1}+(\rho_0\sn\theta-\zeta)\zeta g^{03}]
-\mathcal{L}_{\rm AB}
(1-\zeta^{-1}\rho_0\sn\theta) 
\Bigr\} \varphi_{AB},
 \label{phiAB0}
\eea
where $t_{AB}=t_B-t_A$ and $\varphi_{AB}=\varphi_B-\varphi_A$ are defined. We only consider the case $a<M$,\footnote{We do not consider the case $a>M$ and $a=M$ because the former  leads to a naked singularity and the latter is unstable~\cite{Carroll:2004st}.} so that $a/r_1<r_g/r_1$ holds. For convenience, we assume
\be
O\Bigl(\frac{a}{r_1}\Bigr)\sim O\Bigl(\frac{r_g}{r_1}\Bigr).
\label{para}
\ee
Therefore, expanding \eqref{phiAB0} at the third order in the two quantities (\ref{para}), we get:
\be
\phi_{AB} \approx
\frac{1}{\hbar} \mathcal{E} t_{AB}\Bigl[-\frac{r_g}{2r_1}-\frac{3r_g^2}{8r_1^2}-\frac{5r_g^3}{16r_1^3}+ \frac{a^2 r_g}{2r_1^3} \cos^2(\theta_1)\Bigr]
+\frac{1}{\hbar} \varphi_{AB}
\sin^2(\theta_1) \Bigl(\mathcal{E} \frac{a r_g}{r_1}
 -\frac{1}{2} 
\mathcal{L}_{\rm AB}
  \frac{a^2 r_g}{r_1^3}
 +\mathcal{E} \frac{a r_g^2}{r_1^2}
 \Bigr),
 \label{phiAB}
\ee
where we regard $(r_g/r_1)^i (a/r_1)^j$ as a term of the order $(i+j)$. The quantity $\mathcal{L}$ is given by (see Appendix~\ref{time and angle})
\be
\mathcal{L}=\frac{\mathcal{E}}{g_{00}}(-g_{03} +v^\varphi \Gamma_{33}  \sqrt{g_{00}}),
\label{L given}
\ee
where $v^j$ is the three-dimensional velocity and $\Gamma_{ij}$ is the three-dimensional metric tenor defined by~\cite{Landau:1975pou}\footnote{To distinguish the three-velocity \eqref{v components} from the four-velocity, we emphasize that the velocity $v$ (given by \eqref{vsquare}), which appears in the phase differences, is the ratio between the proper length and the observer's proper time, namely $v=\dd L/\dd\tau$. The first equation in \eqref{v components} is actually equivalent to the definition $v^k=\dd x^k/\dd\tau$ (see Sec.~88 in Ref.~\cite{Landau:1975pou}). 
Pay attention that here the metric is not diagonal. }
\be
v^k=\frac{\dd x^k}{\sqrt{g_{00}} (\dd x^0+\frac{g_{0i}}{g_{00}}
 \dd x^i)},\qquad
\Gamma_{ij} =-g_{ij}+\frac{g_{0i}g_{0j}}{g_{00}}.
\label{v components}
\ee
For the gravitational phases $\phi_{BC}$, $\phi_{AD}$, and $\phi_{DC}$, the derivation is similar to $\phi_{AB}$. Combining these phases, we can get the phase difference between the paths ADC and ABC. Hence, with the following relations (see Appendix~\ref{time and angle}):
\be
r_2\approx r_1 +\frac{l\cos(\gamma)}{\sqrt{-g_{11}(r_1,\theta_1)} },
\qquad
\theta_2\approx \Bigl| \theta_1-
\frac{l\sin(\gamma)}{\sqrt{-g_{22}(r_1,\theta_1)}} \Bigr|,
\label{r2 theta2}
\ee
we can expand the phase difference in the neighborhoods of $r_1$ and $\theta_1$, and for simplicity we only keep the first order terms of $l/r_1$. Then relate the time, the angle, and the energy with the observations (Appendix~\ref{time and angle}):
\bea
&&t_{AB}\approx s\Bigl(
\frac{1}{v\sqrt{g_{00} }}-\frac{g_{03} }{g_{00}\sqrt{\Gamma_{33}} }
\Bigr),
\quad
\varphi_{AB}\approx\frac{s}{\sqrt{\Gamma_{33} }},
\label{t varphi}
\\
&&\mathcal{E}=
\begin{cases}
m(1-v^2)^{-1/2} \sqrt{g_{00}}, &\text{for massive particles},\\
\hbar \omega \sqrt{g_{00}}, &\text{for massless particles}.
\end{cases}
\label{mathcalE lambda}
\eea
With the above steps, we derive
the phase difference between the paths ADC and ABC as follows (see Appendix~\ref{DPD} for more details)
\bea
\delta\phi
&\approx&
\frac{\mathcal{E}_0 l s}{\hbar r_1 } \Bigl\{
\frac{1}{v}\Bigl[
\cos(\gamma)\Bigl(\frac{r_g}{2r_1}+\frac{r_g^2}{2r_1^2}+\frac{r_g^3}{2r_1^3}
+\frac{a^2 r_g}{4r_1^3}\bigl(1-7\cos^2(\theta_1) \bigr)
\Bigr)
+\sin(2\theta_1)
\sin(\gamma)\frac{a^2 r_g }{2r_1^3}
\Bigr]
\nonumber\\
&&
+\frac{a^2 r_g }{r_1^3}\sin(\theta_1)\Bigl[ v \Bigl(
\cos(\gamma)\sin(\theta_1)
+\frac{3}{2}\sin(\gamma)\cos(\theta_1)
\Bigr)
+\frac{1}{2}\sin(\theta_1-\gamma)
\Bigl(v-\sqrt{v^2-(v^r)^2- (r_1 v^\theta)^2}\Bigr)
\Bigr]
\nonumber\\
&&
-\frac{a r_g}{r_1^2}\Bigl(2\cos(\theta_1)\sin(\gamma)+\cos(\gamma)\sin(\theta_1) 
\Bigr)
-\frac{a r_g^2}{r_1^3}
\Bigl(
  \cos(\theta_1)\sin(\gamma)
 +\frac{3}{2}\cos(\gamma)\sin(\theta_1)
 \Bigr)
\Bigr\},
\label{PhaseDifference total}
\eea
where $v^r$ and $v^\theta$ (defind in \eqref{v components}) are the velocity components at the point B corresponding to the path BC, and $\mathcal{E}_0$ is defined by
\be
\mathcal{E}_0=
\begin{cases}
m(1-v^2)^{-1/2}, &\text{for massive particles},\\
\hbar \omega, &\text{for massless particles}.
\end{cases}
\label{mathcalE lambda0}
\ee
By the way, the expression $\sqrt{v^2-(v^r)^2-(r_1 v^\theta)^2}$ in \eqref{PhaseDifference total} can be replaced by $|v\cos(\zeta)|$, where $\zeta$ is a base angle of the parallelogram\footnote{Because this expression only appears in the third order terms in \eqref{PhaseDifference total}, at the point B corresponding to the path BC we have
\bea
\sqrt{v^2-(v^r)^2-(r_1 v^\theta)^2}
\approx r_1\sin(\theta_1) v^\varphi
=r_1\sin(\theta_1) \frac{\dd \varphi}{\dd \tau}
=r_1\sin(\theta_1) \frac{\dd l}{\dd \tau}\frac{\dd L_\varphi}{\dd l}\frac{\dd \varphi}{\dd L_\varphi}
=r_1\sin(\theta_1) v\, |\!\cos(\zeta)| \sqrt{\Gamma_{33}^{-1}}
\approx |v \cos(\zeta)|,
\nonumber
\eea
where \eqref{v varphi} and \eqref{dLsquare} have been used in the first and the last second steps respectively.
}.
In particular, for $\gamma=0$ and $\gamma=\pi/2$, the gravitational phase differences are respectively:
\bea
\delta\phi |_{\gamma=0}
&\approx&
\frac{\mathcal{E}_0 l s}{\hbar r_1 } \Bigl\{
\frac{1}{v}\Bigl[
\frac{r_g}{2r_1}+\frac{r_g^2}{2r_1^2}+\frac{r_g^3}{2r_1^3}
+\frac{a^2 r_g}{4r_1^3}\Bigl(1-7\cos^2(\theta_1) \Bigr)
\Bigr]
\nonumber\\
&&
+\frac{a^2 r_g }{r_1^3}\sin(\theta_1)\Bigl[ v\sin(\theta_1)
+\frac{1}{2}\sin(\theta_1)
\Bigl(v-\sqrt{v^2-(v^r)^2- (r_1 v^\theta)^2}\Bigr)
\Bigr]
\nonumber\\
&&
-\frac{a r_g}{r_1^2}\sin(\theta_1) 
-\frac{3 a r_g^2}{2 r_1^3}\sin(\theta_1)
\Bigr\},
\label{phi pi 0}
%
\\
\delta\phi |_{\gamma=\frac{\pi}{2}}
&\approx& 
\frac{\mathcal{E}_0 l s}{\hbar r_1 } \Bigl\{
\frac{1}{v}
\sin(2\theta_1)
\frac{a^2 r_g }{2r_1^3}
+\frac{a^2 r_g }{r_1^3}\sin(\theta_1)\Bigl[ 
\frac{3}{2} v\cos(\theta_1)
-\frac{1}{2}\cos(\theta_1)
\Bigl(v-\sqrt{v^2-(v^r)^2- (r_1 v^\theta)^2}\Bigr)
\Bigr]
\nonumber\\
&&
-\frac{2a r_g}{r_1^2}\cos(\theta_1) 
-\frac{a r_g^2}{r_1^3}
  \cos(\theta_1)
\Bigr\}.
\label{phi pi over2}
\eea
The prediction \eqref{PhaseDifference total} can be tested experimentally by measuring the fringe shift as a function of $\gamma$. 
As for the non-relativistic particles, \eqref{PhaseDifference total} is reduced to
\bea
\delta\phi^{\rm NR}
&\approx&
\frac{m l s}{\hbar r_1 } \Bigl\{
\Bigl(\frac{1}{v}+\frac{v}{2}\Bigr)\Bigl[
\cos(\gamma)\Bigl(\frac{r_g}{2r_1}+\frac{r_g^2}{2r_1^2}+\frac{r_g^3}{2r_1^3}
+\frac{a^2 r_g}{4r_1^3}\bigl(1-7\cos^2(\theta_1) \bigr)
\Bigr)
+\sin(2\theta_1)
\sin(\gamma)\frac{a^2 r_g }{2r_1^3}
\Bigr]
\nonumber\\
&&
+\frac{a^2 r_g }{r_1^3}\sin(\theta_1)\Bigl[ v \Bigl(
\cos(\gamma)\sin(\theta_1)
+\frac{3}{2}\sin(\gamma)\cos(\theta_1)
\Bigr)
+\frac{1}{2}\sin(\theta_1-\gamma)
\Bigl(v-\sqrt{v^2-(v^r)^2- (r_1 v^\theta)^2}\Bigr)
\Bigr]
\nonumber\\
&&
-\frac{a r_g}{r_1^2}\Bigl(2\cos(\theta_1)\sin(\gamma)+\cos(\gamma)\sin(\theta_1) 
\Bigr)
-\frac{a r_g^2}{r_1^3}
\Bigl(
  \cos(\theta_1)\sin(\gamma)
 +\frac{3}{2}\cos(\gamma)\sin(\theta_1)
 \Bigr)
\Bigr\},
\label{PhaseDifference NR}
\eea
where we have neglected the terms of $O(v^2)$ and higher orders.

From \eqref{PhaseDifference total} we can find that the quantity $a$ only appears in the second and higher order terms. We can also find that in the Newtonian limit the equation \eqref{PhaseDifference total} reproduces the result \eqref{COWpre} on the earth. Indeed, in such limit we have $v\ll 1$ and the phase difference is dominated by the first term in \eqref{PhaseDifference total}, namely
\be
\delta\phi_{\rm mass}
\approx
\frac{m r_g s}{2\hbar r_1^2 v} l\cos(\gamma)
=\frac{m^2  s l\cos(\gamma)\lambda_d }{2\pi\hbar^2} \frac{r_g }{2r_1^2}.
\label{Newtonian limit}
\ee
On the other hand, since $r_g/(2r_1^2)=g$, \eqref{Newtonian limit} is equivalent to \eqref{COWpre} by setting $\gamma=0$. Equation \eqref{Newtonian limit} can also be derived from the gravitational phase directly evaluated in the Newtonian limit:
\be
\phi\approx
\frac{1}{\hbar}\int S_0 \dd t,
\ee
by expanding the result in the ratios $r_g/r$ and $a/r$ and only keeping the first order term.

Notice that \eqref{PhaseDifference total} holds only when the condition $s\ll \sqrt{\Gamma_{33}}$ is satisfied (recall the second equation in \eqref{t varphi}). If the latter condition is violated, the equation \eqref{PhaseDifference total} should be modified. Take $\theta_1=0$ and $\theta_1=\pi$ as examples, then the second equation in \eqref{t varphi} should be replaced by the equation $\varphi_{AB}=\pi$. Correspondingly, \eqref{PhaseDifference total} should be replaced by the following expression (see the last paragraph in the Appendix~\ref{DPD}) 
\bea
\delta\phi |_{\theta_1=0,\pi}=
&\approx&
\frac{\mathcal{E}_0 l s}{\hbar r_1 } \Bigl\{
\frac{1}{v}\Bigl[
\cos(\gamma)\Bigl(\frac{r_g}{2r_1}+\frac{r_g^2}{2r_1^2}+\frac{r_g^3}{2r_1^3}
+\frac{a^2 r_g}{4r_1^3}\bigl(1-7\cos^2(\theta_1) \bigr)
\Bigr)
+\sin(2\theta_1)
\sin(\gamma)\frac{a^2 r_g }{2r_1^3}
\Bigr]
\nonumber\\
&&-\frac{a r_g^2}{2 r_1^3} \cos(\gamma)\sin(\theta_1)
\Bigr\}
+\frac{\pi\mathcal{E}_0 l }{\hbar} \Bigl\{
\frac{a^2 r_g }{r_1^3}v \sin^2(\theta_1)\Bigl(
\cos(\gamma)\sin(\theta_1)
+\frac{3}{2}\sin(\gamma)\cos(\theta_1)
\Bigr)
\nonumber\\
&&
-\frac{a r_g}{r_1^2}\sin(\theta_1) \Bigl(\cos(\theta_1)\sin(\gamma)+\sin(\theta_1+\gamma)
\Bigr)
-\frac{a r_g^2}{r_1^3}
\sin(\theta_1)\sin(\theta_1+\gamma)
\Bigr\}.
\label{PhaseDifference pi}
\eea

\subsection{Impact of the angular momentum of the black hole}

We here discuss the contribution coming from the spin of the black hole. According to \eqref{phiAB} the quantity $a$ only appears in the second and higher order terms (we remind that that the order of $(r_g/r_1)^i (a/r_1)^j$ is $(i+j$).) Furthermore, we claim that in the region $r\gg r_g$ the quantity $a$ does not appear at the first order term in the local gravitational phase (here local means that the path is short enough so that hold $\int S_\beta\dd x^\beta\approx S_\beta\delta x^\beta$, where $\delta x^\beta$ are coordinates' differences). Indeed, we can expand the function $S_\beta$ in \eqref{S beta2} respect to the parameters $\kappa_g$ and $\kappa_a$, where $\kappa_g=r_g/r$ and $\kappa_a=a/r$. Hence, we get:
\be
(S_\beta)\approx
\begin{pmatrix}
\mathcal{E}\Bigl(-\frac{1}{2}\kappa_g-\frac{3}{8} \kappa_g^2-\frac{5}{16} \kappa_g^3+ \frac{1}{2} \cos^2(\theta) \kappa_a^2  \kappa_g 
\Bigr)\\
-\frac{1}{2} P^1 \kappa_g
-\frac{1}{4}  P^2 r \sin(2 \theta )\kappa_a^2
-\frac{5}{8} P^1 \kappa_g^2
-\frac{11}{16} P^1 \kappa_g^3
-\frac{1}{4} P^1  [\cos (2 \theta )-3]\kappa_a^2 \kappa_g
\\
\frac{1}{8}  P^1 r \sin (2 \theta ) (2 \kappa_a^2 +\kappa_a^2 \kappa_g)
\\
 \sin^2(\theta ) \Bigl(\mathcal{E} r \kappa_a \kappa_g
 -\frac{1}{2} \mathcal{L} \kappa_a^2 \kappa_g
 +\mathcal{E} r \kappa_a \kappa_g^2
 \Bigr)
\end{pmatrix},
\label{S beta3}
\ee
where $P^1$ and $P^2$ are given by the following equations~\cite{Chandrasekhar1998}, 
\be
(u^1)^2=E^2\frac{R(r)}{\rho^4},
\qquad
(u^2)^2=E^2\frac{\Theta(\theta)}{\rho^4},
\label{u1square}
\ee
and $R(r)$ and $\Theta(\theta)$ are defined as\footnote{Note that the forms of $R(r)$ and $\Theta(\theta)$ in \eqref{u1square} are different from those in (185) and (186) of Chapter 7 of Ref.~\cite{Chandrasekhar1998}.}
\bea
R(r)&=&r^4 +r^2(a^2-\lambda^2-\eta_0)+2Mr[(a-\lambda)^2+\eta_0]-a^2\eta_0 -\delta_1 \frac{r^2\Delta}{E^2},
\label{R}
\\
\Theta(\theta)&=&\eta_0 +a^2\cos^2\theta -\lambda^2\cot^2\theta -\delta_1\frac{a^2\cos^2\theta}{E^2},
\label{Theta}
\eea
and the parameter $\delta_1$ is defined by
\be
\delta_1=
\begin{cases}
1, \quad\text{for massive particles,}\\
0, \quad\text{for massless particles,}
\end{cases}
\ee
$\lambda=L/E$, $\eta_0=\mathscr{L}/E^2$, and $\mathscr{L}$ is a separation constant in the equations of motion. Then expanding $P^1$ and $P^2$, and plugging them into \eqref{S beta3}, we can find that $\kappa_a$ only appears in the second and higher order terms of $S_\beta$. On the other hand, the local gravitational phase can be written as $\phi\approx S_\beta\delta x^\beta$. Therefore, the quantity $a$ does not appear in the first order terms of the local gravitational phase.

This conclusion is also true for the gravitational phase difference, as \eqref{PhaseDifference total} shows. Therefore, the contribution of the quantity $a$ can be regarded as a small modification to the case of the Schwarzschild spacetime. Theoretically, we can measure the fringe shift between different values of the angle $\theta_1$ to detect the contribution of $a$. However, 
this is not an economic way because we need to move the setup significantly. An alternative way is to rotate the parallelogram along the axis $l$ shown in FIG.~\ref{experimentKerr}. For simplicity, we flip it so that the positions of A and B swap. Correspondingly, the second equation in \eqref{t varphi} should be changed to
\be
\varphi_{AB}\approx
-\frac{s}{\sqrt{\Gamma_{33} }},
\label{varphi2}
\ee
while the expression for $t_{AB}$ is not changed. Besides, we need to make the change $\varphi_{DC}\rightarrow -\varphi_{DC}$ and  reverse the angular momentum such that $\mathcal{L}\rightarrow -\mathcal{L}$. Plugging these changes into \eqref{delta phi a}, \eqref{delta phi b} and \eqref{delta phi c} in the Appendix~\ref{DPD}, we can derive a new phase difference. Let us denote it as $(\delta\varphi)_2$, then the fringe shift for the rotation is
\bea
n&=&\biggl|\frac{(\delta\phi)_2-\delta\phi}{2\pi}\biggr|
\nonumber\\
&=&\biggl|\frac{\mathcal{E}_0 ls}{\pi\hbar r_1}
\Bigl[
\frac{a r_g}{r_1^2}\Bigl(
2\sin(\gamma)\cos(\theta_1)+\cos(\gamma)\sin(\theta_1)
 \Bigr)
+\frac{a r_g^2}{r_1^3}\sin(\gamma+\theta_1) 
\Bigr]
\biggr|,
\label{N pi}
\eea 
where $\delta\phi$ is given in \eqref{PhaseDifference total} and $\mathcal{E}_0$ is defined in \eqref{mathcalE lambda0}. The quantity $n$ shows the impact of the angular momentum of the black hole on the interference. In \eqref{N pi} we can find that the fringe shifts vanish for $a=0$. This is not surprising because flipping the parallelogram along the axis $l$ does not affect the result of the interference in a Schwarzschild spacetime due to spherical symmetry. 

\subsection{Numerical results and discussion}
For simplicity, we here assume $\gamma=0$, we restore the SI units, and use the spin parameter 
\be
a_*=\frac{a c^2}{GM}
=\frac{2a}{r_g}. 
\label{a star}
\ee
As we showed in \eqref{PhaseDifference total} and \eqref{PhaseDifference cc}, the shape of the parallelogram only makes difference at the third order and higher orders in $\delta\phi$. Therefore, in the case $\gamma=0$, for simplicity we assume that the path BC is along the radius, such that $v^r=v$ and $v^\theta=0$ hold in the phase difference \eqref{PhaseDifference total}.

\subsubsection{Massive particles}
In the case of non-relativistic massive particles, the phase difference \eqref{PhaseDifference NR} is:
\be
\delta\phi_{\rm mass}(\gamma=0)
\approx
\delta\phi_1+\delta\phi_2+\delta\phi_3,
\label{gamma0 mass}
\ee
where $\delta\phi_1$, $\delta\phi_2$, and $\delta\phi_3$ are the first, second, and third order terms respectively, namely
\bea
\delta\phi_1&=&\frac{m c l s}{2\hbar r_1}
\Bigl(\frac{mc\lambda_d}{2\pi\hbar}+\frac{\pi\hbar}{mc\lambda_d}\Bigr)
\frac{r_g}{r_1},
\qquad
\delta\phi_2=\frac{m c l s}{2\hbar r_1}\Bigl[\Bigl( \frac{mc\lambda_d}{2\pi\hbar}+\frac{\pi\hbar}{mc\lambda_d}\Bigr)
-a_*\sin(\theta_1)\Bigr]\frac{r_g^2}{r_1^2},
\label{deltaphi12}
\\
\delta\phi_3&=&\frac{m c l s}{2\hbar r_1}
\Bigl\{
\Bigl(\frac{mc\lambda_d}{2\pi\hbar}+\frac{\pi\hbar}{mc\lambda_d}\Bigr)
-\frac{3}{2}a_*\sin(\theta_1)
+\frac{1}{2}a_*^2\Bigl[
\frac{1}{4}\Bigl(1-7\cos^2(\theta_1)\Bigr)
\Bigl(\frac{mc\lambda_d}{2\pi\hbar}+\frac{\pi\hbar}{mc\lambda_d}\Bigr)
\nonumber\\
 &&+\frac{3\pi\hbar}{mc \lambda_d} \sin^2(\theta_1)
\Bigr]
\Bigr\}
\frac{r_g^3}{r_1^3}.
\label{deltaphi3}
\eea
We can find that the phase difference is proportional to the area of the parallelogram. The fringe shift \eqref{N pi}, corresponding to flipping the parallelogram along the axis $l$, is:
\be
n_{\rm mass}
\approx
\biggl|
\frac{mc ls}{2\pi\hbar r_1} a_* \sin(\theta_1)\Bigl(\frac{r_g^2}{r_1^2}+\frac{r_g^3}{r_1^3}
\Bigr)
\biggr|,
\label{nmass0}
\ee
where we have let $\gamma=0$ and neglected the second and higher order terms in $v/c$, and we remind that the Schwarzschild radius is given by $r_g=2GM/c^2$. From \eqref{PhaseDifference total} we can find $\delta\phi(\gamma=\pi)=-\delta\phi(\gamma=0)$. Therefore, if we change the angle $\gamma$ from $0$ to $\pi$, we get a fringe shift
\be
N_{\rm mass}=\Bigl| \frac{\delta\phi_{\rm mass}(\gamma=\pi)-\delta\phi_{\rm mass}(\gamma=0)}{2\pi}\Bigr|
=\Bigl| \frac{\delta\phi_{\rm mass}(\gamma=0)}{\pi}\Bigr|.
\label{defineN}
\ee
In the following we discuss two examples in which the particles that interfere are neutrons. 

(I) The earth as the gravitational source. In this example, we neglect the spin of the earth so that $a_*\approx 0$. For the parameter $r_1$, we assume the equatorial radius of the earth. For the setup of the experiment, we take the parameters in~\cite{Overhauser:1974}, i.e. 
\be
l s=6\times 10^{-4} {\rm m}^2,
\qquad
\lambda_d=1.42\times 10^{-10} {\rm m}.
\label{lambda d}
\ee
And for the constants in \eqref{deltaphi12} and \eqref{deltaphi3}, we use the values given in~\cite{Carroll}. Therefore, we get the results:\footnote{Equations \eqref{deltaphi12} and \eqref{deltaphi3} are used here even though \eqref{para} is violated, because \eqref{phiAB} still holds for the case $a=0$.} 
\be
\delta\phi_1=33.502,
\qquad
\delta\phi_2=4.660\times 10^{-8},
\qquad
\delta\phi_3=6.482\times 10^{-17}.
\label{mass I}
\ee
We can find that the second and the third order terms are much smaller than the first order term. As for the fringe according to \eqref{nmass0} we get:
\be
n_{\rm mass}=0 
\ee
because of $a_*\approx 0$. Combining \eqref{mass I}, \eqref{gamma0 mass}, and \eqref{defineN}, we get:
\be
N_{\rm mass}=10.664.
\label{earthN}
\ee
The value \eqref{earthN} is nearly the same as the result in~\cite{Overhauser:1974}, which agrees with the claim that the equation \eqref{PhaseDifference total} produces the result \eqref{COWpre} on the earth in the Newtonian limit.

(II) The black hole in Cygnus X-1 as the gravitational source. We take the distance between the black hole and the earth to be the value of $r_1$. The parameters are given by~\cite{Miller-Jones:2021plh,Zhao:2021bhj}\footnote{Here we do not consider the uncertainties shown in the references \cite{Miller-Jones:2021plh,Zhao:2021bhj}. Moreover, in such papers the authors do not give the angle $\theta_1$ directly, but give the binary orbital inclination $i=27.51^\circ$. However, as stated in Ref.~\cite{Zhao:2021bhj}, the spin axis of the black hole is assumed to be aligned with the orbital angular momentum. Therefore, the angle $\theta_1$ is equal to the inclination $i$.} 
\be
M=21.2 M_\odot,
\qquad
a_*>0.9985,
\qquad
r_1=2.22 {\rm kpc},
\qquad
\theta_1=27.51^\circ,
\label{parameters X1}
\ee
where $M_\odot$ is the mass of the sun. For simplicity, we assume the value $a_*=0.9985$. For the area of the parallelogram and the wavelength of the neutron, we still use the parameters \eqref{lambda d}. Thus we get:
\be
\delta\phi_1=2.049\times 10^{-18},
\qquad
\delta\phi_2=1.873\times 10^{-33},
\qquad
\delta\phi_3=7.506\times 10^{-49}.
\label{mass II}
\ee
The gravitational phase difference is totally dominated by the first order term. For the fringe shift we get:
\be
n_{\rm mass}=2.557\times 10^{-39},
\label{nmass number}
\ee
when the parallelogram is flipped along the axis $l$. Similar to \eqref{earthN}, we get the fringe shift
\be
N_{\rm mass}=6.524\times 10^{-19},
\label{NII}
\ee
corresponding to changing the angle $\gamma$ from $0$ to $\pi$, which is much smaller than the fringe shift in the example (I). 

\subsubsection{Massless particles}
Similar to \eqref{gamma0 mass}, according to \eqref{PhaseDifference total} for massless particles we have: 
\be
\delta\phi_{\rm massless}(\gamma=0)
\approx
\delta\phi_1+\delta\phi_2+\delta\phi_3,
\label{gamma0 massless}
\ee
where 
\be
\delta\phi_1=\frac{\pi l s }{\lambda_0 r_1}\frac{r_g}{r_1},
\quad
\delta\phi_2=\frac{\pi l s }{\lambda_0 r_1}
\Bigl(1-a_*\sin(\theta_1)\Bigr)\frac{r_g^2}{r_1^2},
\quad
\delta\phi_3=\frac{\pi l s }{\lambda_0 r_1}
\Bigl[1-\frac{3}{2}a_*\sin(\theta_1) 
+\frac{a_*^2}{16}\Bigl(1-13\cos(2\theta_1)
\Bigr)\Bigr]\frac{r_g^3}{r_1^3}.
\label{psi123}
\ee
And the equation \eqref{N pi} is simplified to
\be
n_{\rm massless}
=
\biggl|\frac{ls}{\lambda_0 r_1}\sin(\theta_1)a_* \Bigl(\frac{r_g^2}{r_1^2}+\frac{r_g^3}{r_1^3}\Bigr)
 \biggr|.
\label{n massless0}
\ee
From \eqref{psi123} we can find the phase difference is proportional to the area of the parallelogram and inversely proportional to the wavelength $\lambda_0$ of the particles. As an example, we consider gamma rays and adopt the parameters
\be
l s=6\times 10^{-4} {\rm m}^2,
\qquad
\lambda_0=10^{-12} {\rm m}.
\label{lambda 0}
\ee
Then we repeat the computations in (I) and (II).  

For the example (I) we get:
\bea
&&\delta\phi_1=4.111\times 10^{-7},
\qquad
\delta\phi_2=5.719\times 10^{-16},
\qquad
\delta\phi_3=7.955\times 10^{-25},
\nonumber\\
&&n_{\rm massless}=0,
\qquad
N_{\rm massless}=1.309\times 10^{-7}.
\label{massless I}
\eea
Note that the definition for the fringe shift $N_{\rm massless}$ is similar to \eqref{defineN}. 

For the example (II) we get:
\bea
&&\delta\phi_1=2.515\times 10^{-26},
\qquad
\delta\phi_2=1.239\times 10^{-41},
\qquad
\delta\phi_3=-1.973\times 10^{-57},
\nonumber\\
&&n_{\rm massless}=3.375\times 10^{-42},
\qquad
N_{\rm massless}=8.005\times 10^{-27}.
\label{massless II}
\eea

\subsubsection{Discussion}
Comparing the results in the example (I) with those in the example (II), we find that $N_{\rm I}\gg N_{\rm II}$ holds for both massive and massless particles, where the subscripts denote the two examples. Such inequality is explained by  $N\approx |\delta\phi_1/\pi| \propto r_g/r_1^2$ and $(r_g/r_1^2)_{\rm I}\gg (r_g/r_1^2)_{\rm II}$. Therefore, if we want to increase the fringe shift $N$, we can increase the ratio $r_g/r_1^2$. For example, to let $N_{\rm mass}\approx 1$ in (II), we can decrease the distance to be $r_1\approx 5.533\times 10^{10}{\rm m} \approx 1.793\times10^{-9}{\rm kpc}$ which is much less than the distance $2.22{\rm kpc}$ in \eqref{parameters X1} but still satisfies the condition $r_1\gg r_g\approx 6.262\times 10^4{\rm m}$. In order to increase $N$ we can also increase the area of the parallelogram, according to \eqref{deltaphi12} and \eqref{psi123}. Moreover, for this purpose, in the massive case we can use more massive or slower particles. While for the massless case we can use more energetic particles to increase $N$. As for $n$, to increase its value, we can increase the ratio $r_g^2/r_1^3$, the area of the parallelogram, the spin parameter, or the quantity $\sin(\theta_1)$, according to \eqref{nmass0} and \eqref{n massless0}. For example, in (II) we can decrease the distance to be $r_1\approx 9.368\times10^6{\rm m}$ to let $n_{\rm mass}\approx 1$. Furthermore, we can use more massive particles or more energetic massless particles to increase $n$.

Now we compare the massive case with the massless case. Comparing the values of $N$ in \eqref{earthN} and \eqref{NII} with those in \eqref{massless I} and \eqref{massless II} respectively, we can find that $N_{\rm mass}$ is much greater than $N_{\rm massless}$, although a very small value for $\lambda_0$ is chosen. Moreover, comparing the value of $n$ in \eqref{nmass number} with its value in \eqref{massless II}, we can find $n_{\rm mass}\gg n_{\rm massless}$. Therefore, we conclude that it is more difficult to detect the fringe shifts for massless particles in comparison with the massive case.

Finally, comparing $n_{\rm mass}$ with $N_{\rm mass}$ and comparing $n_{\rm massless}$ with $N_{\rm massless}$ in these examples, we find $n\ll N$ for both cases. This is because $N \approx |\delta\phi_1/\pi|$ is dominated by the first order terms according to \eqref{defineN}, while all the terms of $n$ have orders higher than one according to \eqref{nmass0} and \eqref{n massless0}. Therefore, it is easier to detect $N$ than to detect $n$. Additionally, according to \eqref{deltaphi12} and \eqref{psi123}, the phase difference $\delta\phi_1$ depends on $r_g$, and, according to \eqref{nmass0} and \eqref{n massless0}, the fringe shift $n$ depends on both $a_*$ and $r_g$. Therefore, inversely we can determine the mass of the black hole and its spin parameter according to the measured fringe shifts, following the following steps: 
First we should measure the fringe shift $N$ to determine the mass $M$. For massive particles it is determined by
\be
M=
\frac{2\pi^2 r_1^2 \hbar^2 c^2 \lambda_d}{G ls (m^2 c^2 \lambda_d^2 +2\pi^2\hbar^2)}N_{\rm mass},
\label{M mass}
\ee
while for massless particles it is determined by
\be
M=\frac{c^2 \lambda_0 r_1^2}{2Gls}N_{\rm massless}, 
\label{M massless}
\ee 
then we should measure the fringe shift $n$, from which, given the mass $M$, one can determine the spin parameter $a_*$ as follows. For simplicity we only keep up to second order terms in \eqref{nmass0} and \eqref{n massless0}. Hence, plugging \eqref{M mass} and \eqref{M massless} into these equations, we can determine the spin parameter $a_*$.
For massive particles it is:
\be
a_*=
\frac{ls(m^2 c^2 \lambda_d^2+2\pi^2\hbar^2)^2}{8\pi^3\hbar^3\sin(\theta_1) r_1 mc\lambda_d^2}\frac{n_{\rm mass}}{N_{\rm mass}^2},
\label{a mass}
\ee
while for massless particles it is:
\be
a_*=\frac{ls}{\sin(\theta_1) \lambda_0 r_1}\frac{n_{\rm massless}}{N_{\rm massless}^2}.
\label{a massless}
\ee 

\section{Conclusion\label{Conclusion}}
The gravitational phase difference has been expressed as the integral of a function $S_\beta$ defined by the product of the four-momentum, the metric, and the gauge gravitational potential, which is expressed by the tetrad. 
This is the way to calculate the gravitational phase for a general given spacetime. 
However, as an explicit example, in this paper we considered the case of the Kerr spacetime and we studied a particles' interference experiment (FIG.~\ref{experimentKerr}) analogous to the COW experiment, but in the Kerr spacetime.

We calculated the phase difference for massive and massless particles respectively. 
We found that the angular momentum of the black hole only appears in the second or higher order terms in the phase difference. 
As a generalization, we have proved that the angular momentum density $a$ does not appear in the first order terms of the local gravitational phase at large distance respect to the Schwarzschild radius, namely for $r\gg r_g$.  Then we have evaluated the fringe shifts for several examples, compared the results, and discussed how to increase the fringe shifts. Concretely, in order to increase the fringe shifts, we should take a larger black hole's mass, decrease the distance from it, or increase the area of the parallelogram. For this purpose, we could also choose more massive and slower particles or more energetic massless particles. According to the numerical results, we found that it is more difficult to measure the fringe shifts for massless particles than those for massive particles. In the end, we showed how to determine the mass of the black hole and its spin parameter by the measurement of the fringe shifts.  

We here propose some potential extensions of our work. Besides the interference with paths along a parallelogram we could consider other configurations, or, in addition to the asymptotically flat region, we could consider the region closer to the black hole gravitational radius. Additionally, the numerical examples should not be limited to the black hole in Cygnus X-1, but other examples should be discussed in the future. Finally, considering the universality of the gravitational phase \eqref{phi_g}, we could apply it to other spacetimes. For example, we could consider other compact objects such as binary black holes, neutron stars, and rotating galaxies or black holes beyond Einstein's theory of gravity \cite{Giacchini:2021pmr,Nicolini:2008aj,Modesto:2010uh,Modesto:2010rv}.

Finally, recalling that the phases derived in this paper are based on Teleparallel Gravity, it is essential to compare our results with those in general relativity. Even though the equation of motion in the former is equivalent to the one in the latter (see Sec.~\ref{Role of B}), the quantum aspects of these theories are not necessary the same. Now let us compare the non-relativistic phase difference \eqref{PhaseDifference NR} with the one obtained in general relativity. In Ref.~\cite{Wajima:1996cu}, the authors calculated the phase difference of a quantum interferometer experiments on the earth, with the rotation of the earth taken into account. In the weak field limit and up to the first order in the post-Newtonian approximation, they found
\bea
\delta\phi^{\rm PN}
&=&
\frac{m^2 g A \lambda}{2\pi\hbar^2}\sin\mu
+\frac{2m}{\hbar}\vec{\omega}\cdot\vec{A}
+\frac{2m}{5\hbar}\frac{r_g}{R}\Bigl[ \vec{\omega} -3\Bigl(\frac{\vec{R}}{R}\cdot\vec{\omega}\Bigr)\frac{\vec{R}}{R}
\Bigr]
\cdot\vec{A}
\nonumber\\
&&-\frac{1}{2} \frac{r_g}{R}
\Bigl(\frac{m^2 g A \lambda}{2\pi\hbar^2}\sin\mu\Bigr)
+\frac{3}{2}\Big(\frac{\lambda_C}{\lambda}\Bigr)^2
\Bigl(\frac{m^2 g A \lambda}{2\pi\hbar^2}\sin\mu\Bigr),
\label{terms PN}
\eea
where $g$ is the gravitational acceleration, $\vec{A}$ is the area vector enclosed by the interferometry loop, $\mu$ is the angle between $\vec{A}$ and the position vector $\vec{R}$ of the interferometer, $R$ is the radius of the earth, $\lambda$ and $\lambda_C$ are the de Broglie wavelength and the Compton wavelength respectively, and $\vec{\omega}$ is the angular velocity vector of the earth with its magnitude related with the Kerr parameter by $a=2R^2\omega/5$. For simplicity, we use $\delta\alpha_j$ to denote the $j^{\rm th}$ term in \eqref{terms PN}, where $j=1,2,...,5$. According to Ref.~\cite{Wajima:1996cu},  these terms are interpreted as follows: The first term $\delta\alpha_1$ is just the result predicted in Ref.~\cite{Overhauser:1974}, verified by the COW experiment~\cite{Colella:1975dq}; the term $\delta\alpha_2$ due to Sagnac effect~\cite{Sagnac1,Sagnac2} is caused by the rotation of the interferometer (recall that this experiment is on the earth); the term $\delta\alpha_3$ is due to the Lense-Thirring effect~\cite{Lense}; finally, the terms $\delta\alpha_4$ and $\delta\alpha_5$ correspond to the redshift corrections to the potential energy and the kinetic energy respectively. To compare the result~\eqref{PhaseDifference NR} with \eqref{terms PN}, we need to rewrite the latter according to the parameters in our result. Notice that the second term in \eqref{terms PN} is absent here, namely $\delta\alpha_2=0$, because in FIG.~\ref{experimentKerr} the interferometer is assumed to be not rotating. Then according to the relation between $a$ and $\omega$, and the equations:
\be
g=\frac{r_g}{2R^2},
\qquad
\mu=\frac{\pi}{2}-\gamma', \qquad
\vec{R}\cdot\vec{\omega}= R\omega\cos\theta',
\qquad
\vec{R}\cdot\vec{A}= RA\cos(\mu),
\qquad
\vec{\omega}\cdot\vec{A}= \omega A\cos(\theta'+\mu),
\ee
where $\gamma'$ is the angle between $\vec{R}$ and the plane of the interferometry, we can rewrite the remaining terms in \eqref{terms PN} as follows:
\bea
&&\delta\alpha_1=\frac{m A \cos(\gamma')}{2\hbar R}\frac{1}{v}\frac{r_g}{R},
\qquad
\delta\alpha_3=-\frac{mA}{\hbar R}\Bigl(2\cos(\theta')\sin(\gamma')+\cos(\gamma')\sin(\theta')\Bigr)\frac{a r_g}{R^2},
\nonumber\\
&&\delta\alpha_4=-\frac{m A \cos(\gamma')}{4\hbar R}\frac{1}{v}\frac{r_g^2}{R^2},
\qquad
\delta\alpha_5=
\frac{3 m A \cos(\gamma')}{4\hbar R}\frac{r_g v}{R},
\label{terms PN2}
\eea
where the units $c=1$ has been used. To compare our result with \eqref{terms PN2}, we neglect the terms of the third order such that the phase difference \eqref{PhaseDifference NR} reduces to
\bea
\delta\phi^{\rm NR}&=&\frac{m ls \cos(\gamma)}{2\hbar r_1 }\frac{1}{v}\frac{r_g}{r_1}
-\frac{mls}{\hbar r_1}\Bigl(2\cos(\theta_1)\sin(\gamma)+\cos(\gamma)\sin(\theta_1)\Bigr)
\frac{a r_g}{r_1^2}
\nonumber\\
&&+\frac{m ls \cos(\gamma)}{2\hbar r_1 }\frac{1}{v}\frac{r_g^2}{r_1^2}
+\frac{m ls \cos(\gamma)}{4\hbar r_1}\frac{r_g v}{r_1}
+\frac{m ls \cos(\gamma)}{4\hbar r_1}\frac{1}{v}\frac{r_g^2 v^2}{r_1^2}.
\label{NR second}
\eea
The first two terms in \eqref{NR second} coincide with those in \eqref{terms PN2}, while the third and fourth terms are different from those of \eqref{terms PN2} only in the coefficients, and the last term in \eqref{NR second} can be neglected  compared with other terms because $(r_g v/r_1)^2$ is very small. We notice that the area $A$ in Ref.~\cite{Wajima:1996cu} is defined in flat space (see (3.20) in Ref.~\cite{Wajima:1996cu}), while in this paper the area is defined by the length in curved spacetime (see \eqref{dLsquare}). However, even using the later, the form of the phase differences in \eqref{terms PN2} are not changed (see the last paragraph in Appendix~\ref{time and angle}), such that the above conclusions do not change. Finally, we mention that the first two terms in \eqref{NR second} also coincide with the result in Ref.~\cite{Kuroiwa}, where the authors study the same experiment on the earth. They use an approximation to the first order of $M$ and $a$, and neglect the terms of $O(v^2)$. 

Therefore, in the weak field limit, the non-relativistic phase difference \eqref{PhaseDifference NR} based on the theory of Teleparallel Gravity, reproduces partly the result of post-Newtonian approximation in general relativity. In particular, reproduces the result in the COW experiment and the term of the Lense-Thirring effect. It looks a little strange that the predictions from Teleparallel Gravity in the interference experiment are not exactly the same as those from general relativity. Indeed, consider that the equation of motions for a particle in Teleparallel Gravity is identical to the geodesic equation in general relativity (as mentioned in Sec.~\ref{Role of B}). We have to admit that we do not know how to explain such difference, and we simply notice that the derivations for the phase in this paper and in Ref.~\cite{Wajima:1996cu} are different. In Ref.~\cite{Wajima:1996cu}, the phase is found by constructing the quantum Hamiltonian of a non-relativistic particle in the weak gravitational field up to the first order of the post-Newtonian approximation, and plugging the Hamiltonian into the Schr\"odinger equation. While in our paper, following Ref.~\cite{Aldrovandi:2013wha}, the phase is constructed by separating a gauge potential $B{^a}_\mu$ analogous to the electromagnetic potential from the Lagrangian of the particle (see the last paragraph in Sec.~\ref{Role of B}). Given that we have not provided a wave equation satisfied by \eqref{GP}, this phase is a conjecture to some extent. In spite of this, we think that it is reasonable in the perspective of the analogy with electromagnetism, and since it successfully reproduces the result of COW experiment. However, the gravitational phase in Teleparallel Gravity deserves more investigations before to be able to help us revealing the differences between Teleparall Gravity and general relativity in such quantum aspects\footnote{As suggested by the referee of this paper, if the Aharonov-Bohm effect is sensitive to the potential ${B^a}_\mu$, it could provide a
possible way to distinguish general relativity from Teleparallel Gravity.}.

\acknowledgments
This work was supported by the Basic Research Program of the Science, Technology, and Innovation Commission of Shenzhen Municipality (grant no. JCYJ20180302174206969). 

\appendix 
\section{Derivations for some formulas\label{time and angle}}
In this appendix we derive some equations used in Sec.~\ref{interference}. Let us prove \eqref{L given} firstly. According to \eqref{conserved quantities} and recalling $g_{30}=g_{03}$ in Kerr spacetime, we get
\be
\mathcal{L}=-\mathcal{E}\Bigl(g_{03}\frac{\dd t}{\dd\varphi} +g_{33} \Bigr)
\Bigl( g_{00}\frac{\dd t}{\dd\varphi} +g_{03} \Bigr)^{-1}.
\label{L matric}
\ee
The expression for $\dd t/\dd\varphi$ is found by letting $k=3$ in the definition of the velocity \eqref{v components}, namely 
\be
\frac{\dd t}{\dd \varphi}
=\frac{1}{v^\varphi \sqrt{g_{00}} }-\frac{g_{03}}{g_{00}}.
\label{dtdphi}
\ee 
Hence, replacing \eqref{dtdphi} into \eqref{L matric}, we prove \eqref{L given}.

Now we derive \eqref{r2 theta2}. As shown in~\cite{Landau:1975pou}, in a spacetime with its metric independent on the time coordinate, a distance $L$ is defined as the integral of the distance element $\dd L$ given by:
\be
\dd L^2 =\Gamma_{ij}\dd x^i \dd x^j,
\label{dLsquare}
\ee
where the three-dimensional metric tensor $\Gamma_{ij}$ is defined in \eqref{v components}. Applying \eqref{dLsquare} to the radial component of $l$ and the component perpendicular to the radius (see FIG.~\ref{experimentKerr}), we find:
\bea
&&l\cos(\gamma) =\sqrt{\Gamma_{11}} \dd r
\approx \sqrt{-g_{11}} (r_2-r_1),
\\
&&l \sin(\gamma) =\bigl |\sqrt{\Gamma_{22}} \dd \theta \bigr|
\approx
\begin{cases}
 \sqrt{-g_{22}} (\theta_1-\theta_2),\quad &\text{when $l\sin(\gamma)\sqrt{-g_{22}^{-1}}\le  \theta_1$,}\\
  \sqrt{-g_{22}} (\theta_1+\theta_2)
  \quad &\text{when $l\sin(\gamma)\sqrt{-g_{22}^{-1}}>  \theta_1$,}
 \end{cases}
\eea
which imply the two equations in \eqref{r2 theta2} respectively.

Now we show how to derive \eqref{t varphi}. We know that $\dd r\approx 0$ and $\dd\theta\approx 0$ on the path AB (assumption (a) in Sec.~\ref{TPre}), therefore, using \eqref{dLsquare} to this path, we have the following relation: 
\be
\dd\varphi\approx\frac{\dd L}{\sqrt{\Gamma_{33} }}.
\label{dvarphi 1}
\ee
Hence, combining \eqref{dvarphi 1} with \eqref{dtdphi}, we find the relation between time coordinate and length, i.e., 
\be
\dd t \approx \Bigl(
\frac{1}{v^\varphi \sqrt{g_{00}} }-\frac{g_{03}}{g_{00}}
\Bigr)
\frac{\dd L}{\sqrt{\Gamma_{33} }}.
\label{ds 1}
\ee
Integrating \eqref{dvarphi 1} and \eqref{ds 1}, and taking \eqref{v varphi 0} into account, we derive the two equations in \eqref{t varphi}. 

As for the energy of a massive particle in \eqref{mathcalE lambda}, we take directly the result from Ref.~\cite{Landau:1975pou} (see Sec.~88 in~\cite{Landau:1975pou} ). While for a massless particle, its energy reads~\cite{Carroll:2004st}:
\be
\mathcal{E}= V \hbar \omega,
\label{EV}
\ee
where $\omega$ is its frequency measured by a static observer, while $V$ is the redshift factor given by:\footnote{In the metric signature $(-,+,+,+)$ the redshift factor is replaced by $V=\sqrt{-K_\mu K^\mu}$.}
\be
V=\sqrt{K_\mu K^\mu},
\label{redshift factor}
\ee
where $K^\mu$ is the Killing vector related to the time-translation invariance. Here by a static observer we mean that the four-velocity of the observer is proportional to the Killing vector~\cite{Carroll:2004st}. Inserting the Killing vector $K^\mu=(1,0,0,0)$ into \eqref{redshift factor}, we obtain $V=\sqrt{g_{00}}$. Finally, plugging this result into \eqref{EV}, we obtain the second expression in \eqref{mathcalE lambda}. 

Now we prove the statement in Sec.~\ref{Conclusion} that the forms of the terms in \eqref{terms PN2} do not change when we use the area defined in the Kerr spacetime to re-express them. Firstly, the metric (2.1) in Ref.~\cite{Wajima:1996cu} can be rewritten as\footnote{We believe that there is a typo in the last term of $g_{00}$ in (2.1) of Ref.~\cite{Wajima:1996cu}. Here we have made a modification.}
\be
\dd s^2 =\Bigl[ 1+2\Phi +2\Phi^2 -\Bigl(\frac{r_g a}{r'^2}\Bigr)^2 \sin^2(\theta')
\Bigr]\dd t'^2 
+\frac{2r_g a}{r'}\sin^2(\theta')\dd \varphi'\dd t'
-(1-2\Phi) 
\bigl(\dd r'^2 +r'^2\dd \theta'^2 +r'^2\sin^2(\theta')\dd \varphi'^2
\bigr),
\label{ds square}
\ee
where $\Phi=-r_g/(2r')$ is the Newtonian potential, and the coordinates $(t', r', \theta',\varphi')$ relate with the asymptotically static coordinates $(t, x', y',z')$ by 
\be
t'=t,\qquad
x'=r'\sin(\theta')\cos(\varphi'),\qquad
y'=r'\sin(\theta')\sin(\varphi'),\qquad
z'=r'\cos(\theta').
\ee
The terms in \eqref{terms PN2} are derived by the following integral~\cite{Wajima:1996cu}
\be
\delta\alpha_j
=-\frac{1}{\hbar}\oint H_j \dd t,
\ee
where the loop encloses the interferometer, and $H_j$ are defined by
\be
H_1=m\Phi,
\qquad
H_3=\frac{2 r_g R^2}{5 r'^3}\vec{\omega}\cdot \vec{J},
\qquad
H_4=\frac{m}{2}\Phi^2,
\qquad
H_5=\frac{3\Phi \vec{p}\,^2}{2m},
\ee
where $\vec{J}=\vec{r}\times \vec{p}$ is the angular momentum defined in flat space, and $\vec{p}=m\vec{v}$. For simplicity, assume that the loop of the interferometer is a parallelogram. As we mentioned in Sec.~\ref{Conclusion}, the area $A$ in \eqref{terms PN2} is defined in flat space~\cite{Wajima:1996cu}. If we take a new area defined by the length in the Kerr spacetime (see \eqref{dLsquare}), the term $\delta\alpha_1$ now reads
\be
\delta\alpha_1
=-\frac{m r_g}{2\hbar} \Bigl(\frac{1}{r'_2}-\frac{1}{r'_1} \Bigr) t_{AB}
=-\frac{m r_g}{2\hbar} \Bigl(\frac{1}{r'_1+l\cos(\gamma)/\sqrt{-g_{11} }}-\frac{1}{r'_1} \Bigr) s\Bigl( \frac{1}{v\sqrt{g_{00}}}-\frac{g_{03}}{g_{00}\sqrt{\Gamma_{33}}}
\Bigr)
\approx
\frac{m l s \cos(\gamma)}{2\hbar r'_1}\frac{1}{v}\frac{r_g}{r'_1},
\label{alpha1other}
\ee
where we have used \eqref{r2 theta2} and \eqref{t varphi} (they still hold in the coordinates $(t', r', \theta',\varphi')$), and have neglected the third order and higher orders terms of $r_g$ and $a$. Here the product $ls$ is the area of the interferometry loop, and $l$ and $s$ are lengths defined in the Kerr spacetime. Similar calculations lead to
\be
\delta\alpha_4\approx
-\frac{m l s \cos(\gamma)}{4\hbar r'_1}\frac{1}{v}\frac{r_g^2}{{r'_1}^2},
\qquad
\delta\alpha_5\approx
\frac{3 m ls \cos(\gamma)}{4\hbar r'_1}\frac{r_g v}{r'_1}.
\label{delta4delta5}
\ee
As for $\delta\alpha_3$, it is still given by the result of Ref.~\cite{Wajima:1996cu},
\be
\delta\alpha_3=-\frac{mA}{\hbar r'_1}\Bigl(2\cos(\theta'_1)\sin(\gamma')+\cos(\gamma')\sin(\theta'_1)\Bigr)\frac{a r_g}{{r'_1}^2},
\label{delta33}
\ee
where $A=l' s'$ is the area defined in flat space. We need to re-express $\delta\alpha_3$ according to the area $l s$. For this purpose, applying \eqref{dLsquare} to the parallelogram in the radial direction, we obtain
\be
l\cos(\gamma) = L_{r'}\approx \sqrt{\Gamma_{11}}\,   l'\cos(\gamma'),
\qquad
\Rightarrow\,
l'\cos(\gamma') \approx \frac{l\cos(\gamma)}{\sqrt{\Gamma_{11}} }.
\label{lprime1}
\ee
Similarly, in the direction of $\theta'$ and $\varphi'$ we get respectively
\be
l'\sin(\gamma') 
\approx \frac{r'_1}{\sqrt{\Gamma_{22}}} l\sin(\gamma),
\qquad
s'\approx \frac{r'_1 \sin(\theta'_1)}{\sqrt{\Gamma_{33}}}s .
\label{lprime2}
\ee
Replacing \eqref{lprime1} and \eqref{lprime2} into \eqref{delta33}, expanding the expression, and neglecting the third order and higher order terms, we get
\be
\delta\alpha_3
\approx
-\frac{m l s}{\hbar r'_1}\Bigl(2\cos(\theta'_1)\sin(\gamma)+\cos(\gamma)\sin(\theta'_1)\Bigr)\frac{a r_g}{{r'_1}^2}.
\label{delta333}
\ee
Finally, comparing \eqref{alpha1other}, \eqref{delta4delta5} and \eqref{delta333} with \eqref{terms PN2}, we can see that the forms of $\delta\alpha_j$ are not changed.

\section{Derivations for the phase difference\label{DPD}}

In this appendix, we show how to derive the phase difference \eqref{PhaseDifference total} between the paths ADC and ABC in FIG.~\ref{experimentKerr}. Lest us start writing the coordinates of the points A, B, C, and D as follows:
\be
{\rm A}(t_A, r_1, \theta_1, \varphi_A),
\quad
{\rm B}(t_A+t_{AB}, r_1, \theta_1, \varphi_A+\varphi_{AB}),
\quad
{\rm C}(t_D+t_{DC}, r_2, \theta_2, \varphi_D+\varphi_{DC}),
\quad
{\rm D}(t_D, r_2, \theta_2, \varphi_D),
\label{coordinatesABCD}
\ee
where we defined $t_{AB}=t_B-t_A$,  $\varphi_{AB}=\varphi_B-\varphi_A$, $t_{DC}=t_C-t_D$, and $\varphi_{DC}=\varphi_C-\varphi_D$. 

Hence, we write down the phases of each path in the following way. As we mentioned in the assumption (a) in Sec.~\ref{TPre}, we have $\dd r\approx 0$ and $\dd\theta\approx 0$ on the paths AB and DC. Moreover, according to \eqref{S beta2}, we know that all the components $S_\beta$ are independent on $t$ and $\varphi$.\footnote{As for $P^1$ and  $P^2$ which appear in the expressions of $S_1$ and $S_2$, they are also independent of $t$ and $\varphi$ (see~\eqref{u1square}).} 
Therefore, \eqref{PhiK} simplifies to:
\bea
\phi_{AB}&\approx& \frac{1}{\hbar} (S_0^A t_{AB} +S_3^A \varphi_{AB})_{AB},
\label{phaseAB}
\\
\phi_{DC}&\approx& \frac{1}{\hbar} (S_0^D t_{DC} +S_3^D \varphi_{DC})_{DC},\label{phaseDC}
\eea
where the superscripts A and D denote the positions, and the subscripts AB and DC denote the paths. As for the path AD, we have
\bea
\phi_{AD}
&=&
\frac{1}{\hbar}\Bigl(\int S_\beta \dd x^\beta
\Bigr)_{AD} 
\nonumber\\
&=&
\frac{1}{\hbar}\Bigl[
S_0 (\vec{r}_a) t_{AD}
+ S_1 (\vec{r}_b) r_{AD}+ S_2 (\vec{r}_c) \theta_{AD}+ S_3(\vec{r}_d) \varphi_{AD}
\Bigr]_{AD}
\nonumber\\
&=&
\frac{1}{\hbar}\Bigl\{\!
\bigl[ S_0 (\vec{r}_a) \!-\! S_0^A \!+\! S_0^A \bigr] t_{AD}
+ \bigl[ S_1 (\vec{r}_b) \!-\! S_1^A   \!+\! S_1^A
\bigr] r_{AD}
+ \bigl[ S_2 (\vec{r}_c)
 \!-\! S_2^A \!+\! S_2^A \bigr] \theta_{AD}
+\bigl[ S_3(\vec{r}_d) \!-\! S_3^A \!+\! S_3^A
\bigr]\varphi_{AD}
\!
\Bigr\}_{AD}
\nonumber\\
&\approx& \frac{1}{\hbar} \bigl[ S_0^A (t_D-t_A)
+S_1^A (r_2-r_1) +S_2^A (\theta_2-\theta_1) +S_3^A (\varphi_D-\varphi_A) 
\bigr]_{AD},
\label{phaseAD}
\eea
where we have used the {\em mean value theorem for integrals} in the second step, and $\vec{r}_a$, $\vec{r}_b$, $\vec{r}_c$ and $\vec{r}_d$, are points on AD. The last step in \eqref{phaseAD} holds because both $(S_\beta(\vec{r}_p)-S_\beta^A)_{AD}$ and $x^\beta_{AD}$ are smaller than or equal to $O(l/r_1)$.\footnote{For any point $p$ on the path AD, we have $|r_p-r_1|\le |r_2-r_1|$ and $|\theta_p-\theta_1|\le |\theta_2-\theta_1|$. Therefore, expanding $(S_\beta(\vec{r}_p)-S_\beta^A)_{AD}$ in the neighborhoods of $r_1$ and $\theta_1$, we can find it is smaller than or equal to $O(l/r_1)$. As for $x^\beta_{AD}$, both $r_{AD}$ and $\theta_{AD}$ are $O(l/r_1)$, according to \eqref{r2 theta2}. Finally, for $t_{AD}$ and $\varphi_{AD}$, the geodesic equations in Kerr spacetime imply~\cite{Chandrasekhar1998}
\be
t=\int T_1(r) \dd r
+\int T_2(\theta)\dd \theta,
\qquad
\varphi =\int \Phi_1(r) \dd r
+\int \Phi_2(\theta)\dd \theta,
\nonumber
\ee
where
\be
T_1(r)=\frac{r^2(r^2+a^2)+2Ma r(a-\lambda)}{\Delta\sqrt{R(r)}},
\qquad
T_2(\theta)=\frac{a^2 \cos^2\theta}{\sqrt{\Theta(\theta)}},
\qquad
\Phi_1(r)=\frac{a(r^2+a^2-a\lambda)}{\Delta\sqrt{R(r)}},
\qquad
\Phi_2(\theta)=\frac{\lambda {\rm cosec}^2(\theta)-a}{\sqrt{\Theta(\theta)}},
\nonumber
\ee
with $R(r)$ and $\Theta(\theta)$ defined in \eqref{R} and \eqref{Theta}. Hence, from the {\em mean value theorem for integrals} we have:
\be
t_{AD}=T_1(r_e) r_{AD} +T_2(\theta_f) \theta_{AD},
\qquad
\varphi_{AD}=\Phi_1(r_g) r_{AD} +\Phi_2(\theta_h) \theta_{AD},
\nonumber
\ee
where the subscript $e$, $f$, $g$, and $h$ denote points on AD. Therefore, $t_{AD}$ and $\varphi_{AD}$ 
are also $O(l/r_1)$.}
(Recall the sentence after \eqref{r2 theta2}.) 
Similar to \eqref{phaseAD}, we can derive
\bea
\phi_{BC}&\approx& \frac{1}{\hbar} \bigl[ S_0^B (t_D-t_A +t_{DC}-t_{AB})
+S_1^B (r_2-r_1) +S_2^B (\theta_2-\theta_1) +S_3^B (\varphi_D-\varphi_A +\varphi_{DC}-\varphi_{AB} ) 
\bigr]_{BC},
\label{phaseBC}
\eea
where we have used \eqref{coordinatesABCD}. In order to simplify the expression for $(\phi_{AD}-\phi_{BC})$ we have to show the equality $(S_\beta^A)_{AD}=(S_\beta^B)_{BC}$. We know that $(g_{\mu\nu}^A)_{AD}=(g_{\mu\nu}^B)_{BC}^A$ holds because the metric is independent of $t$ and $\varphi$. Therefore, taking into account the latter conclusion and the expression of $S_\beta$ in \eqref{S beta2}, we have to prove the following equalities, 
\be
(\mathcal{E}^A)_{AD}=(\mathcal{E}^B)_{BC},
\qquad
(\mathcal{L}^A)_{AD}=(\mathcal{L}^B)_{BC},
\qquad
(P^1)^A_{AD}=(P^1)^B_{BC},
\qquad
(P^2)^A_{AD}=(P^2)^B_{BC}.
\label{proof}
\ee
The first equation in \eqref{proof} holds because of the assumption~(b) in Sec.~\ref{TPre}. This assumption also implies $(\vec{v}_A)_{AD}=(\vec{v}_B)_{BC}$ for a the particle on the parallelogram. Thus combining this conclusion with the expression~\eqref{L given}, the second equation in \eqref{proof} is proved. Finally, according to \eqref{u1square}, the last two equations in \eqref{proof} also hold. Since we have proved \eqref{proof}, the relation $(S_\beta^A)_{AD}=(S_\beta^B)_{BC}$ holds. The latter equality together with \eqref{phaseAD} and \eqref{phaseBC} implies:
\be
\phi_{AD}-\phi_{BC}=-\frac{1}{\hbar}\bigl[
S_0^B (t_{DC}-t_{AB})
+S_3^B (\varphi_{DC}-\varphi_{AB} )
\bigr]_{BC}.
\label{ADBC}
\ee
Merging together \eqref{ADBC}, \eqref{phaseAB}, and \eqref{phaseDC}, the phase difference between the paths ADC and ABC reads:
\bea
\delta\phi
&=&\phi_{ADC}-\phi_{ABC}
\nonumber\\
&=&\phi_{DC}-\phi_{AB}+\phi_{AD}-\phi_{BC}
\nonumber\\
&=& \frac{1}{\hbar}\biggl\{
\bigl[ (S_0^D)_{DC} -(S_0^A)_{AB} \bigr] t_{AB} 
+\bigl[ (S_3^D)_{DC} -(S_3^A)_{AB} \bigr] \varphi_{AB}
\nonumber\\
&&
+\bigl[(S_0^D)_{DC} -(S_0^B)_{BC} \bigr] (t_{DC}-t_{AB})
+\bigl[(S_3^D)_{DC} -(S_3^B)_{BC} \bigr] (\varphi_{DC}-\varphi_{AB})
\biggr\}.
\label{delta phi0}
\eea
Recalling the expression of $S_\beta$ in \eqref{S beta2}, we find that $S_0$ does not depend on $\mathcal{L}$, such that $S_0$ keeps its value when the probe particle turns direction at B and D. Hence, we have $(S_0^B)_{BC} = (S_0^B)_{AB}=(S_0^A)_{AB}$. Therefore, taking \eqref{phaseAB} into account, we can rewrite \eqref{delta phi0} as:
\be
\delta\phi=\delta\phi_a+\delta\phi_b +\delta\phi_c,
\label{delta phi01}
\ee
where
\bea
\delta\phi_a
&=& 
\frac{1}{\hbar}\bigl[ (S_0^D)_{DC} t_{AB}+ (S_3^D)_{DC}\varphi_{AB}\bigr] 
-\phi_{AB},
\label{delta phi a}
\\
\delta\phi_b
&=& 
\frac{1}{\hbar}
\bigl[(S_0^D)_{DC} -(S_0^A)_{AB} \bigr] (t_{DC}-t_{AB}),
\label{delta phi b}
\\
\delta\phi_c
&=& 
\frac{1}{\hbar}\bigl[(S_3^D)_{DC} -(S_3^B)_{BC} \bigr] (\varphi_{DC}-\varphi_{AB}).
\label{delta phi c}
\eea
In the following, we compute explicitly the above phase differences. We remind that the phase $\phi_{AB}$ was obtained in \eqref{phiAB}. For the first term in \eqref{delta phi a}, comparing it with \eqref{phaseAB}, we only need to replace $r_1$ with $r_2$, $\theta_1$ with $\theta_2$, and $\mathcal{L}_{AB}$ with $\mathcal{L}_{DC}$ in \eqref{phiAB}. Therefore, \eqref{delta phi a} is equivalent to:
\be
\delta\phi_a=\phi_{AB}(r_1\rightarrow r_2, \theta_1\rightarrow \theta_2, \mathcal{L}_{AB}\rightarrow \mathcal{L}_{DC})
-\phi_{AB}(r_1, \theta_1, \mathcal{L}_{AB}).
\label{delta a1}
\ee
Now we focus on the expression of $\mathcal{L}_{AB}$. According to \eqref{L given}, we need to find the expression for $v^\varphi$. By means of the equation~\cite{Landau:1975pou}:
\be
v^2=\Gamma_{ij} v^i v^j,
\label{vsquare}
\ee
in Kerr spacetime we find:
\be
|v^\varphi |=\sqrt{\frac{v^2+g_{11} (v^r)^2 +g_{22} (v^\theta)^2}{\Gamma_{33}} }.
\label{v varphi}
\ee
On the path AB we have $v^\varphi>0$, $v^r=0$, and $v^\theta=0$, therefore, \eqref{v varphi} simplifies to: 
 \be
v^\varphi=\sqrt{\frac{v^2}{\Gamma_{33}} }.
\label{v varphi 0}
\ee
Replacing \eqref{v varphi 0} into \eqref{L given}, we obtain:
\be
\mathcal{L}_{AB}=\frac{\mathcal{E}}{g_{00}}\Bigl(
-g_{03}+\sqrt{g_{00} \Gamma_{33} v^2}
\Bigr)
\approx
\mathcal{E} r_1\sin(\theta_1) \epsilon,
\label{LAB}
\ee
where $\epsilon$ is given by:
\be
\epsilon=
\begin{cases}
\sqrt{1-\frac{m^2}{\mathcal{E}^2}}, &\text{for massive particles},\\
1, &\text{for massless particles}.
\end{cases}
\label{epsilon}
\ee
We have neglected the terms of $r_g$ and $a$ in the last expression of \eqref{LAB} because, according to \eqref{phiAB}, $\mathcal{L}_{AB}$ only appears in the third order and higher order terms of $\phi_{AB}$. Moreover, we have used \eqref{mathcalE lambda} and expression \eqref{metric} in the last step of \eqref{LAB}. As for $\mathcal{L}_{DC}$, we only need to replace $r_1$ with $r_2$, and $\theta_1$ with $\theta_2$ in~\eqref{LAB}. 
As for \eqref{delta a1}, using \eqref{r2 theta2} we expand $\delta\phi_a$ in the neighborhoods of $r_1$ and $\theta_1$ up to the first order. Finally, we get:
\bea
\delta\phi_a
&\approx&
\frac{\mathcal{E} l}{\hbar r_1} \Bigl\{
t_{AB}\Bigl[
\cos(\gamma)\Bigl(\frac{r_g}{2r_1}+\frac{r_g^2}{2r_1^2}+\frac{r_g^3}{2r_1^3}
+\frac{a^2 r_g}{4r_1^3}\bigl(1-7\cos^2(\theta_1) \bigr)
\Bigr)
+\sin(2\theta_1)
\sin(\gamma)\frac{a^2 r_g }{2 r_1^3}
\Bigr]
\nonumber\\
&&
+\varphi_{AB}\Bigl[
\epsilon\frac{a^2 r_g }{r_1^2}\sin(\theta_1)\Bigl(
\cos(\gamma)\sin^2(\theta_1)
+\frac{3}{4}\sin(\gamma)\sin(2\theta_1)
\Bigr)
\nonumber\\
&&
-\Bigl(
\cos(\gamma)\sin^2(\theta_1) \Bigl(\frac{a r_g}{r_1}
  +\frac{3 a r_g^2}{2 r_1^2}
 \Bigr)
+\sin(\gamma)\sin(2\theta_1) \Bigl(
\frac{a r_g}{r_1}
 +\frac{a r_g^2}{r_1^2}
 \Bigr) 
 \Bigr)
\Bigr]
\Bigr\}.
\label{PhaseDifference a}
\eea
Plugging \eqref{epsilon}, \eqref{t varphi}, and \eqref{mathcalE lambda} into \eqref{PhaseDifference a}, and expanding the resulting expression, we get:
\bea
\delta\phi_a
&\approx&
\frac{\mathcal{E}_0 l s}{\hbar r_1 } \Bigl\{
\frac{1}{v}\Bigl[
\cos(\gamma)\Bigl(\frac{r_g}{2r_1}+\frac{r_g^2}{2r_1^2}+\frac{r_g^3}{2r_1^3}
+\frac{a^2 r_g}{4r_1^3}\bigl(1-7\cos^2(\theta_1) \bigr)
\Bigr)
+\sin(2\theta_1)
\sin(\gamma)\frac{a^2 r_g }{2r_1^3}
\Bigr]
\nonumber\\
&&
+v \frac{a^2 r_g }{r_1^3}\sin(\theta_1)\Bigl(
\cos(\gamma)\sin(\theta_1)
+\frac{3}{2}\sin(\gamma)\cos(\theta_1)
\Bigr)
\nonumber\\
&&
-\frac{a r_g}{r_1^2}\Bigl(2\cos(\theta_1)\sin(\gamma)+\cos(\gamma)\sin(\theta_1) 
\Bigr)
-\frac{a r_g^2}{r_1^3}
\Bigl(
  \cos(\theta_1)\sin(\gamma)
 +\frac{3}{2}\cos(\gamma)\sin(\theta_1)
 \Bigr)
\Bigr\},
\label{PhaseDifference aa}
\eea
where $\mathcal{E}_0$ is defined by:
\be
\mathcal{E}_0=
\begin{cases}
m(1-v^2)^{-1/2}, &\text{for massive particles},\\
\hbar \omega, &\text{for massless particles}.
\end{cases}
\label{mathcalE lambda00}
\ee

As for $\delta\phi_b$ in \eqref{delta phi b}, it is $O(l^2/r_1^2)$ and higher orders
 because both $[(S_0^D)_{DC} -(S_0^A)_{AB}]$ and $(t_{DC}-t_{AB})$ are of the order of $O(l/r_1)$ and higher orders, according to \eqref{r2 theta2}. Therefore, in our approximation (we remind the reader the sentence after \eqref{r2 theta2}), this phase difference is negligible:
\be
\delta\phi_b\approx 0.
\label{PhaseDifference bb}
\ee

As for $\delta\phi_c$ in \eqref{delta phi c}, repeating the above calculations, we find:
\be
\delta\phi_c
\approx
\frac{\mathcal{E}_0 ls }{2\hbar r_1}
\frac{a^2 r_g}{r_1^3} \sin(\theta_1)
\sin(\theta_1-\gamma)
\Bigl(v-\sqrt{v^2-(v^r)^2- (r_1 v^\theta)^2}\Bigr),
\label{PhaseDifference cc}
\ee
where $v^r$ and $v^\theta$ are the components of the velocity at the point B corresponding to the path BC (see the definition~\eqref{v components}).  
In \eqref{PhaseDifference cc} we can find that the phase difference $\delta\phi_c$ is sensible to the direction of the velocity at B along the path BC, therefore, $\delta\phi_c$ depends on the shape of the parallelogram. 

According to the above results, $\delta\phi_b$ and $\delta\phi_c$ are much smaller than $\delta\phi_a$. This is what we expected because according to \eqref{delta phi b} and \eqref{delta phi c}, $\delta\phi_b$ and $\delta\phi_c$ are due to the differences between $t_{AB}$ and $t_{DC}$, and $\varphi_{AB}$ and $\varphi_{DC}$ respectively. These differences are very small compared with $t_{AB}$ and $\varphi_{AB}$, thus, $\delta\phi_b$ and $\delta\phi_c$ can be regarded as small modifications to the phase difference. Merging together \eqref{PhaseDifference aa}, \eqref{PhaseDifference bb}, and \eqref{PhaseDifference cc}, we finally obtain the phase difference~\eqref{PhaseDifference total}. 

Now we show briefly how \eqref{PhaseDifference pi} is derived for $\theta_1=0$ and $\theta_1=\pi$. As for $\delta\phi_a$, we only need to insert $\phi_{AB}=\pi$ into \eqref{PhaseDifference a}, and to repeat the above calculations. Hence, $\delta\phi_b\approx 0$ still holds. Finally, $\delta\phi_c$ can be also neglected, because when $\theta_1=0$ or $\theta_1=\pi$ hold, plugging \eqref{S beta3} into \eqref{delta phi c}, it results:
\be
\delta\phi_c
\approx
\frac{1}{\hbar}\sin^2(\theta_2)
\Bigl(\mathcal{E}\frac{a r_g}{ r_2} 
 -\frac{1}{2} \mathcal{L}_{\rm DC} \frac{a^2 r_g}{r_2^3}
 +\mathcal{E} \frac{a r_g^2}{r_2^2}
 \Bigr)
 (\varphi_{DC}-\varphi_{AB}).
\ee
Here $\sin^2(\theta_2)$ is of the order of $O(l^2/r_1^2)$ and higher orders, hence $\delta\phi_c$ is negligible. Based on above results, we can derive \eqref{PhaseDifference pi}.


{}

\end{document}